\RequirePackage{fix-cm}
\documentclass[smallextended]{svjour3}       % onecolumn (second format)
\smartqed  % flush right qed marks, e.g. at end of proof
\pdfoutput = 1 
\usepackage{graphicx}
\usepackage{multirow}
\usepackage{lscape}
\usepackage{colortbl}
\usepackage{epsfig}
\journalname{European Actuarial Journal}
\usepackage{amssymb}
\journalname{European Actuarial Journal}

\begin{document}

{}

\title{Nonparametric estimation of multivariate distribution function
for truncated and censored lifetime data
}

\titlerunning{Estimation of multivariate distribution for truncated and censored  lifetime data}        % if too long for running head

\author{ Valery Baskakov         \and
        Anna Bartunova 
        %etc.
}


\institute{International Actuarial Advisory Company (IAAC), Llc \at
              Malaya Filevskaya st. 50-80, Moscow, 121433, Russia 
           \and   
           V. Baskakov \at
              Tel.: +7-903-100-2660\\
              \email{chief@actuaries.ru}           %  \\
           \and
           A. Bartunova \at
              Tel.: +44-7376-183291\\
              \email{BartunovaAnna@gmail.com} 
            \and
           % This is a pre-print of an article published in European Actuarial Journal.\\ 
           % The final authenticated version is available online at:\\ \webhttps{https://doi.org/10.1007/s13385-019-00194-1}   
}

\date{Received: 03-05-2017 / Accepted: date}

\maketitle
%\
\begin{abstract}
\footnote{This is a pre-print of an article published in European Actuarial Journal. The final authenticated version is available online at: https://doi.org/10.1007/s13385-019-00194-1}

A number of models for generating statistical data in various fields of insurance, including life insurance, pensions, and general insurance have been considered. It is shown that the insurance statistics data, as a rule, are truncated and censored, and often multivariate. We propose a non-parametric estimation of the distribution function for multivariate truncated-censored data in the form of a quasi-empirical distribution and a simple iterative algorithm for its construction. To check the accuracy of the proposed evaluation of the distribution function for truncated-censored data, simulation studies have been conducted, which showed its high efficiency. The proposed estimates have been tested for many years by the IAAC Group of Companies in the actuarial valuation of corporate social liabilities according to IAS 19 Employee Benefits. Apart from insurance, some results of the work can be used, for example in medicine, biology, demography, mathematical theory of reliability, etc.
\end{abstract}

\section{Introduction}
\label{intro}

%%%%%%%%%%%%%%%%%%%%%%%%%%%%%%%%%%%%%

Indicators of loss and  level of risk associated with an insurance policy are of the greatest interest for an insurance actuary. In general, these indicators are multidimensional random variables defined by their distribution function. Therefore, \ estimating this distribution function or  their parameters \ (e.g., moments, quantiles etc.) is one of the most important and often the most difficult part of an actuarial work. The complexity of this estimation work is mainly due to the complexity of the structure of insurance statistics and underdevelopment of  statistical analysis methods. 

Since insurance is always a time dependent process, one of the indicators of level of risk is naturally time itself. This complicates significantly the statistical data structure  due to the fact that time data are often censored or truncated (e.g., Baskakov et al. \cite{Baskakov2010}, \cite{Baskakova2014}   and many other authors). 

There are a lot of studies handling the problem of non-parametric estimation of the truncated and censored data distribution function. Most of them deal with univariate data such as lifetime. Kaplan and Meier ~\cite{Kaplan1958} proposed a non-parametric estimate of the distribution function based on right-censored data, and Lynden-Bell~\cite{LyndenBell1971} considered a similar estimate for left-truncated data. Later, Tsai~\cite{Tsai1987} followed by Lay and Ying~\cite{Lai1991} modified the Kaplan-Meier estimator for the case of left-truncated and right-censored data. In their works, Peto~\cite{Peto1973}, Turnbull~\cite{Turnbull1976}, Efron and Petrosian~\cite{Efron1999} also discussed methods for constructing non-parametric estimation of the distribution function in the presence of censoring and/or truncation of observations for sampling schemes other than  left-truncated and right-censored. For example, Turnbull~\cite{Turnbull1976} obtains the non-parametric maximum likelihood estimator of the cumulative distribution function from grouped, interval-censored and/or truncated data. Frydman~\cite{Frydman1994} modified this estimate, which allowed applying it for virtually any univariate truncated-censored data scheme.  

Bivariate lifetime models under truncation and/or censoring in the available literature are very limited. 
Most studies are essentially based on parametric or semiparametric models. For example, Frees et al.~\cite{Frees1996}, Carriere~\cite{Carriere2000}, Wang~\cite{Wang2003} and Luciano et al.~\cite{Luciano2008} consider copula models for describing the bivariate distribution function. Dai and Bao~\cite{Dai2009}, Dai et al.~\cite{Dai2016}, proposed an algorithm allowing the use of various parametric functions to transform bivariate data into
univariate data, and evaluating the univariate distribution function using the Kaplan-Meier estimator and its inverse transformation into bivariate distribution. Here, the estimate of the bivariate distribution function is ambiguous, depending on the type of function used to convert data.  

Non-parametric estimates mainly refer to the case when components of the bivariate vector are only right-censored and/or left-truncated. For instance,  Campbell~\cite{Campbell1981}, Van Der Laan~\cite{VanDerLaan1996}, Akritas and Van Keilegom~\cite{Akritas2003} proposed bivariate distribution function estimation for censored data, while  G$\ddot{u}$rler~\cite{Gurler1996} and~\cite{Gurler1997} for truncated data. The similar estimations when both components are censored and truncated have been recently developed by Sankaran and Antony~\cite{Sankaran2007}, Shen and Yan~\cite{Shen2008}, Lopez~\cite{Lopez2012} and Shen~\cite{Shen2014}. The above estimates and their algorithms have been developed for certain truncating and/or censoring schemes of the components of a bivariate vector. Therefore, their adaptation to larger vectors and other truncation and censoring schemes is impossible or difficult. 

In this regard, the work of Baskakov~\cite{Baskakov1996} shall be considered, which deals with the non-parametric evaluation of the multivariate distribution function based on censored data of almost any data type and proposes a simple iterative algorithm for its construction. However, this estimate fails to extrapolate any truncated observations, which somewhat limits its application in actuarial practice. 

In this paper, we employ Baskakov's idea ~\cite{Baskakov1996} and generalize his estimate to the case of multivariate truncated-censored data. The work is organized as follows. Section 2 considers specific examples of truncated-censored data from the field of insurance. Section 3 provides a formal description of the data censorship and truncation process and sets out the purpose of the work. Section 4 provides a non-parametric estimate of the distribution function based on multivariate truncated censored data and a generalized iterative algorithm for its construction. Section 5 describes in detail a non-parametric estimate of the distribution function based on univariate truncated-censored data. Section 6 proposes a simulation study considering the accuracy of the estimate proposed in the univariate case and provides a comparison with known analogues. In Section 7, the proposed estimate is applied to solve a practical actuarial problem based on a real set of bivariate truncated-censored data.

\section{Examples of truncated and censored data}
\label{sec:1}

\paragraph{Example 1. Classical life insurance.} Suppose a $t$-aged person purchase a life insurance policy in  year $y$. An insurance company includes this person's information in its database and starts monitoring his/her life expectancy from that moment. Therefore, the insurance company observes a conditional random variable $x$ – life expectancy of the insured at the age of $t$, i.e., provided that his age $x>t$. If we consider the entire company portfolio, it is obvious that such situation arises for each of the insured based on his/her age at the time of the purchase of the policy. Thus, the insurance company database is a collection of information about conditional random variables with random condition (age of policy purchase is in general a random value)

\begin{equation}
\label{equ:1}
\{x=x_k \mid x>t_k \},   k=1,\ldots,n,
\end{equation}
with $n$ being the number of elements in the insurer database. 

\begin{figure}
\center
  \includegraphics[scale=1.0]{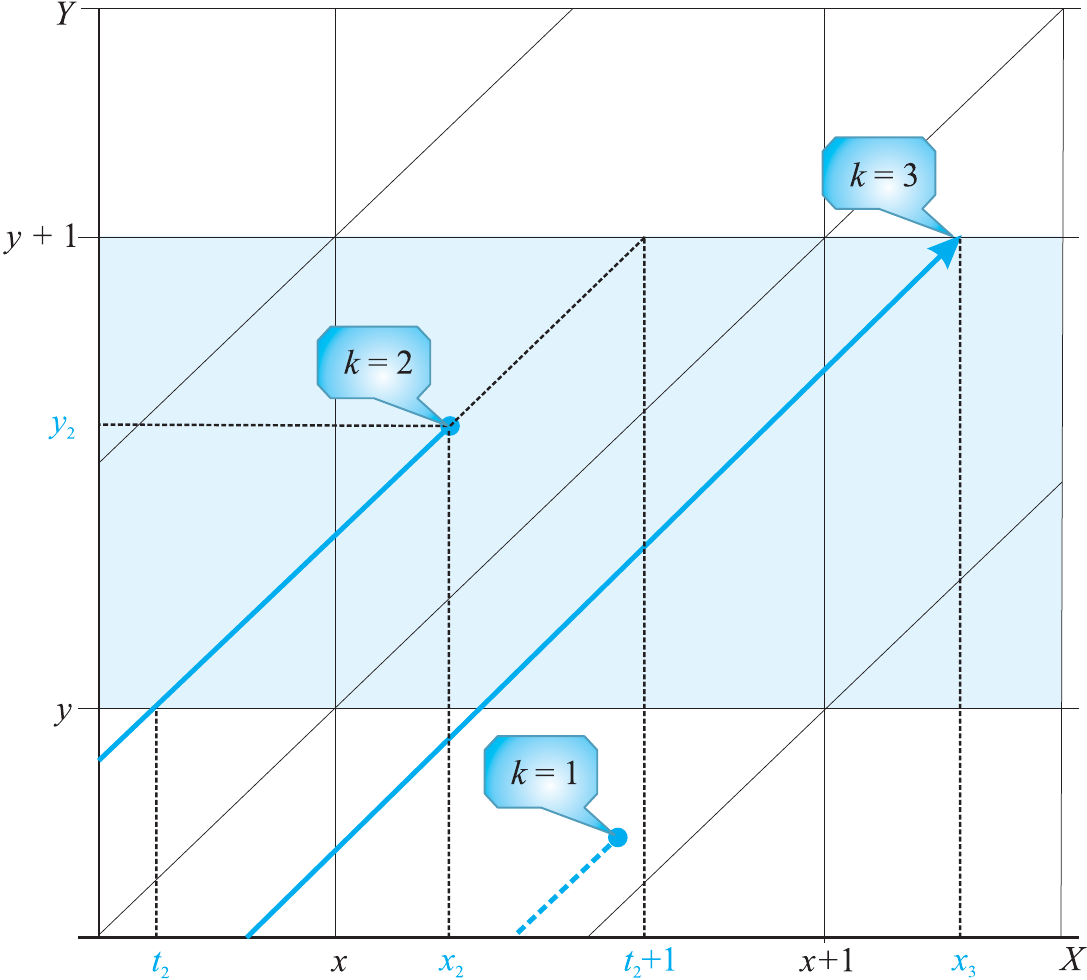}
\caption{The scheme of development of left-truncated and right-censored data}
\label{fig:1}       % Give a unique label
\end{figure}

Such type of data as opposed to an unconditional random variable $x$  is called truncated data. This name is justified by the fact that compared to the traditional statistical sampling
\begin{equation}
\label{equ:2}
\{ x=x_k \},   k=1,\ldots,N
\end{equation}
where $N$ is the sample size, in a truncated sample~(\ref{equ:1}) the value $n$ value cannot be interpreted as statistical sampling size in the conventional sense, since the truncated sample of $n$ was obtained from the total sample of $N$ ($N>n$). Indeed, in this data collection scheme an observer is unable to access the information about events that has happened before the truncation moment $t_k$. It is illustrated in Fig.~\ref{fig:1} showing the Lexis diagram (e.g. Winsch et al.~\cite{Winsch2002}). The truncated data of the type considered with truncation set $T_k=(-\infty,t_k)$ is called left-truncated.

Unfortunately, these are not the only challenge that an actuary faces during insurance statistics analysis. Additional problems are associated with various reasons for termination of the observations of an individual at the age of  $\tau_k$. Some of them are as follows:

	1. The claim occurrence, i.e., death of an individual;
	
	2. The death of an individual due to a cause not being an insurance event (e.g., suicide or death while practicing extreme sports);
	
	3. The cancellation of an insurance contract;
	
	4. The need for actuarial calculations during the period when an individual is alive, etc.

Note that only in the first case the realization of a random variable is fully observed, which corresponds to the data collection in classical mathematical statistics. In all other cases, the insurer knows only that the insured event will happen/would have happened later, if the observation did not stop, and that during the observation from the age $t_k$ to the age $\tau_k$ the claim has not occurred. With $x_k$  as the age at event occurence (in this case the death of the individual) the observer knows only up to a set $C_k \ni x_k$, such data are called censored data. The censored data case considered above, with $C_k=(\tau_k, \infty)$, is called right-censored (the information right from the censoring age  $x_k$ is as if being removed/censored). Thus, the considered statistical information type refers to the samples randomly left-truncated and right-censored. 

In fact, a truncated and/or censored data is very typical to insurance data. Insurance data always cover a time interval and accompanied by a loss (possibly zero loss). Regardless of the insurance type and statistics collection process, the time indicator is always censored (just because actuarial calculations are carried out periodically, when not all the contracts are completed and, therefore, not all the insured events have occurred) and as shown above, the time indicator may be further truncated. Similarly, the value of loss amount also suffers from the same problem, with the only difference being the loss amount not always censored and/or truncated. Let us consider the following example.

\paragraph{Example 2. Motor insurance.} Suppose a policyholder in a motor insurance contract limit the insurance sum to the $u_c$ value equal, for example, to 50-100\% of the vehicle cost, and/or use the franchise  $u_t$ (from 0 to 40\% of the sum insured). Similarly, with example 1, it is clear that $u_c$ insurance sum acts as a censoring value, while the $u_t$ franchise acts as the truncation value. As a result, we again have left-truncated and right-censored sample.

The most studied cases of the truncated-censored data are left-truncated and right-censored samples, which are the subject of many publications, for example, the fundamental work by Klein and Moeschberger~\cite{Klein2003}.The other  schemes of truncated and censored data appear in the literature quite rarely. For example, the International Database of Longevity contains double truncated data, which Gampe~\cite{Gampe2010} used in the study of human mortality beyond age 110.  However, the data available to an actuary to address the most interesting and practically important tasks are, as a rule, multidimensional and have a complex structure. Let us consider some relevant examples.

\paragraph{Example 3. Lifetime joint annuity.} This annuity is issued to two individuals (usually spouses), and paid fully or partially until the death of both policyholders. With spouses generally living together, having common budget and, therefore, the same income, their way of life and living conditions being often similar, there is a reason to believe that their life expectancies are positively correlated random variables. In this regard, to calculate their joint annuity, you should use the joint distribution function of the spouses' lifetime.

The statistical data structure concerning each spouse's life expectancy that is available to an insurance company coincides with the one that the company observes in classical life insurance (see Example 1). In the bivariate case the situation changes fundamentally, and a censored set $C_k$ and a truncated set $T_k$ are more complicated due to their higher dimension:
\begin{equation}
\label{equ:3}
C_k =\left\{
    \begin{array}{cl}
         &  (m>m_k;w=w_k), \ \mbox{with}\ m_k=y_k+t_k;\ y_k>w_k-\tau_k;   \\
         &  (m=m_k;w>w_k), \ \mbox{with}\ w_k=y_k+\tau_k;\ y_k>m_k-t_k;\\
         &  (m=m_k;w=w_k), \ \mbox{with}\ m_k-t_k<y_k; \ w_k-\tau_k<y_k;     \\
         &  (m>m_k;w>w_k), \ \mbox{with}\ m_k-t_k= w_k-\tau_k=y_k           \\
    \end{array}  
\right. 
\end{equation}
and
\begin{equation}
\label{equ:4}
T_k=(m<t_k; w< \infty) \cup (m< \infty;w<\tau_k ), \  k=1,\ldots,n
\end{equation}
with 	$m$ and $w$ the current age of husband and wife respectively;

$t_k$, $m_k$ and  $\tau_k$, $w_k$ the age of truncation (of the insurance contract conclusion), the age of observation (censoring) termination due to death of husband and wife respectively or other reasons;

$y_k$ the time since the contract conclusion until the date as of which data is collected;

$n$ the number of policies sold.

\begin{figure}
\center
  \includegraphics[scale=1.0]{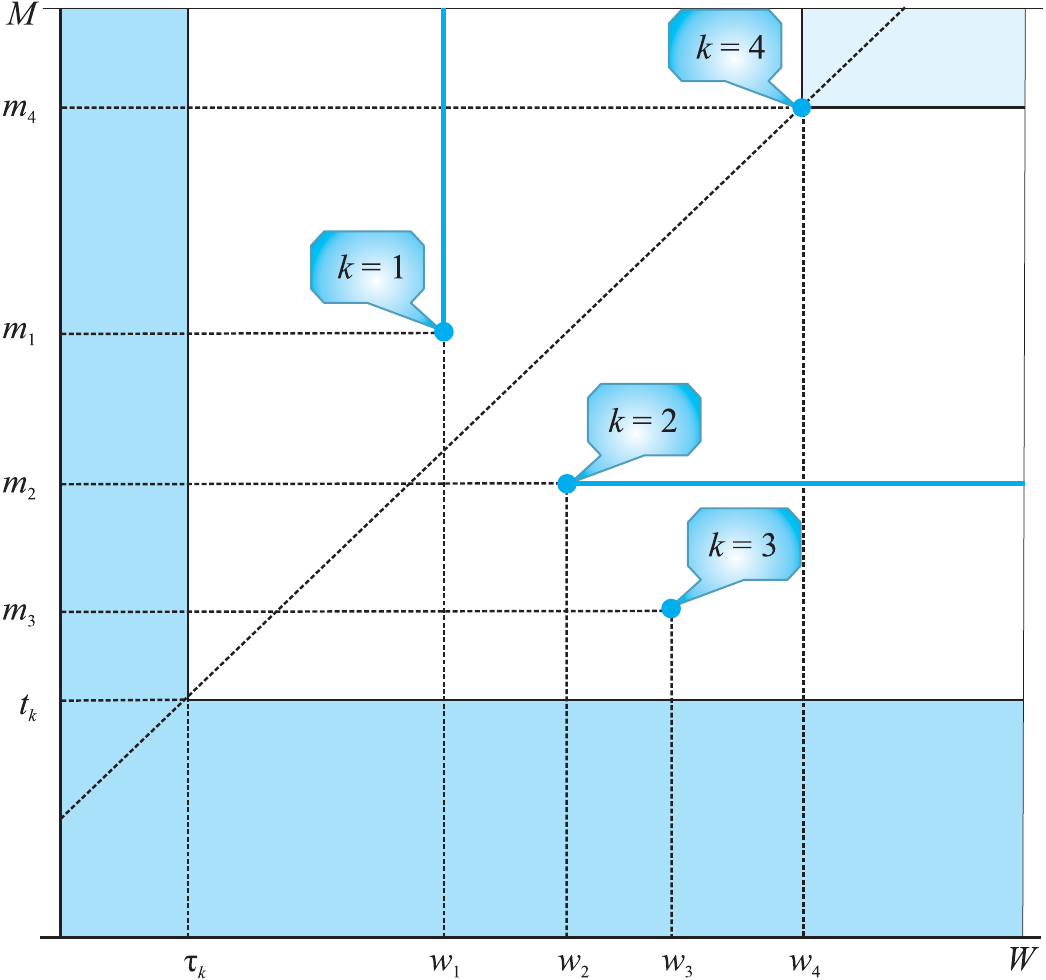}
\caption{Censoring $C_k$ and truncating $T_k$ of a bivariate $(M,W)$ vector set}
\label{fig:2}      
\end{figure}

To visualize possible sets $C_k$ and $T_k$ are represented on Fig.~\ref{fig:2}. To simplify we put that  $t_1= ,\ldots,=t_4$ and $\tau_1= ,\ldots,=\tau_4$, i.e., $T_k$ sets are the same\footnote{This suggestion does not affect information value of the chart since  $T_k$ set is defined by formula ~(\ref{equ:4}) and with accuracy to the coordinates of the  $(t_k,\tau_k)$  point is the same for all the policies sold.}  for all four policies, and the relationship between the ages of censoring and spouses' death are selected in such a way as to show the diversity of $C_k$ sets, with $k$ corresponding to the number of rows in~(\ref{equ:3}).

% ЗДЕСЬ ДОБАВИТЬ, ЧТО ДАННАЯ СХЕМА ШИРОКО ИССЛЕДУЕТСЯ В ЛИТЕРАТУРЕ. ЭТО ЕДИНСТВЕННАЯ СХЕМА

Most studies on the estimation of bivariate distribution from truncated and censored data are somehow connected with the solution of this problem (see the previously mentioned articles, as well as the monograph by Hougaard~\cite{Hougaard2001}, papers by Dabrowska \cite{Dabrowska1988}, Gijbels and G$\ddot{u}$rler~\cite{Gijbels1998}, 
Pruitt~\cite{Pruitt1993} et al.). However, insurance data are more diverse. The following  examples  illustrate this diversity.

\paragraph{Example 4. Selection mortality tables. } In general, selection mortality table are based on the conditional distribution function, e.g., distribution function of life expectancy of individuals with pure endowment or with term life contracts, or having a disability, etc. For this distribution, the selection condition is a categorical variable, and the mortality tables construction is reduced to the problem discussed in Example 1. The only difference is that the statistical sample is based on the selection. However, a quantitative variable can often be used as a selection condition. In the case of disability, such a condition is the age at which a person may become disabled. In this case, it is convenient to talk about the joint distribution of the lifetime $X$ and the age of disability $I$, where marginal distributions at any fixed age of disability $i$ will form selection mortality tables.

\begin{figure}
\center
  \includegraphics[scale=1.0]{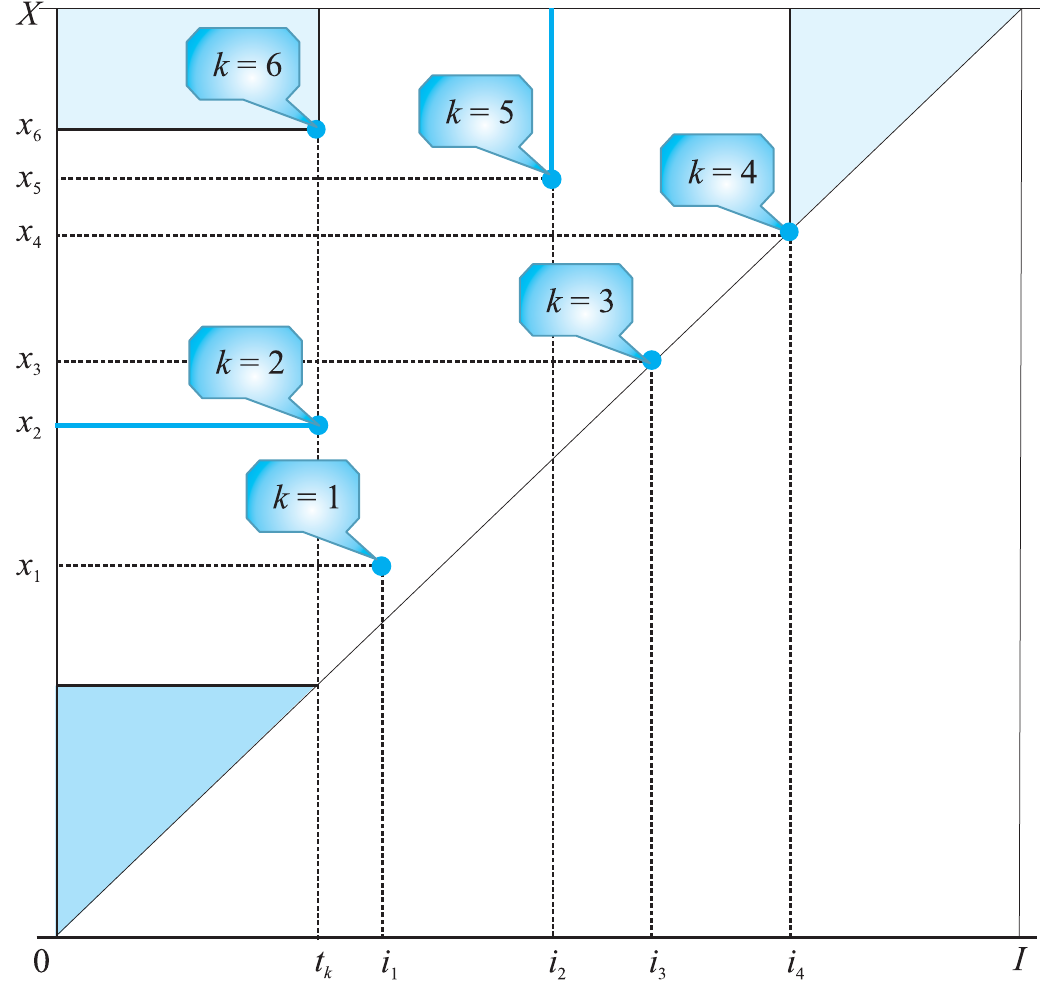}
\caption{Censoring $C_k$ and truncating $T_k$ sets of the bivariate vector $(X,I)$  }
\label{fig:3}       % Give a unique label
\end{figure}

Fig.~\ref{fig:3} shows the structure of statistical information on the random vector $(X,I)$ available for insurance companies. The scope of this vector has a specific triangular representation type and is determined by the system of equations: $X>0$, $I>0$ and $X\ge I$, where equality is achieved in the case of death of a non-disabled persons $(k=3)$. In case of death of a disabled $(k=1)$ person there is also a complete realization of the $(X,I)$ vector, but with, $X>I$. Note that in general, the observations are truncated and censored. Data truncation coincides with the beginning of the observation of an individual
at age $t_k$, and censoring occurs as the result of an observation termination for reasons not associated with his death $(k = 4,$ $5$ or $6)$ and/or when the observation
concerns a person being already disabled and the age of disability is unknown
$(k = 2$ or $6)$.

The figure below shows possible options of censored sets  $C_k$ and a truncated set $T_k$, which, for clarity, are the same for all six policies. Meanwhile the expressions for sets $C_k$ and $T_k$ will not be written explicitly, assuming that a keen reader can do it himself.

\paragraph{Example 5. Selection table of employees' decrement.} In the evaluation of social liabilities in accordance with IAS 19 Employee benefits, some social benefits are often connected with an employee's age and others with his/her years of service. Traditionally the age of an employee is the basis for building mortality and retirement tables and the years of service is used to build table of decrement due to resignation. This approach complicates the analysis, as different bases of decrement tables do not allow to build a multiple decrement table consistent with them directly, without additional assumptions. An attempt to use age as the basis to build all these tables leads to the loss of accuracy of estimations of financial flows linked to the employee's years of service.

The way to cope with this problem is almost obvious. When assessing the enterprise's social liabilities it is necessary to use selection  tables of decrement due to resignation where the basis is an employee's age, and the selection condition is his age at the time of recruitment. The complexity of the available statistical information and the underdeveloped methods of statistical analysis determine the obvious problems of such an approach.

\begin{figure}
\center
  \includegraphics[scale=1.0]{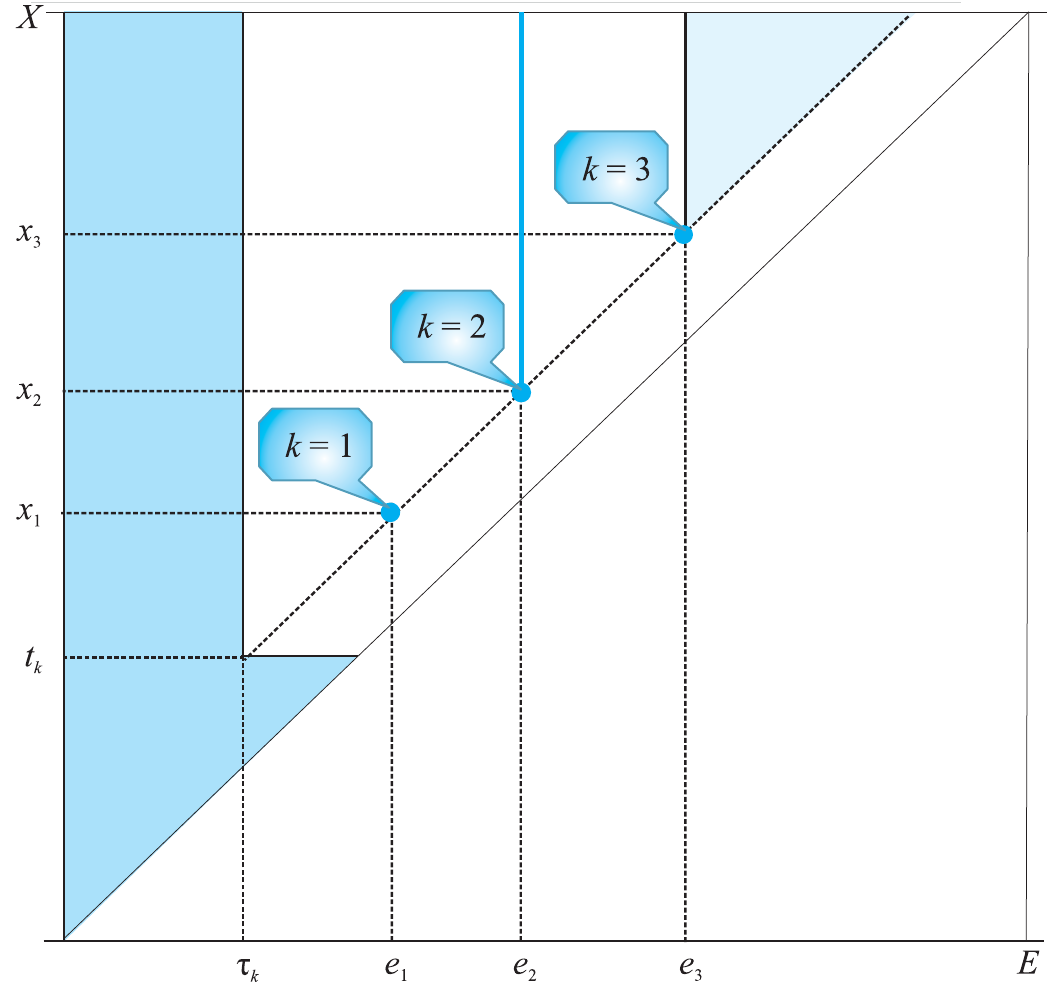}
\caption{Censoring $C_k$ and truncating $T_k$ sets of the bivariate  $(X,E)$ vector}
\label{fig:4}      
\end{figure}

A standard scheme of turnover data collection  is as follows. Human resources department provides the employees' data according to the payroll at the beginning and the end of the reporting period, including date of birth, date of recruitment, date and reason for employment termination. Usually the following reasons for job termination are mentioned: resignation, retirement, death and others. The structure of such statistical information are shown in Fig.~\ref{fig:4}. You may notice that it is similar (but not identical) to the data structure considered in the previous example. In this case, a truncation set is more sophisticated 

\begin{equation}
\label{equ:5}
T_k=(x<t_k  ,e<x) \cup (x \ge t_k  ,e<\tau_k )
\end{equation}
and a possible type of censoring sets is less diverse

\begin{equation}
\label{equ:6}
C_k =\left\{
    \begin{array}{cl}
         &  (x_k, e_k),\ \mbox{with}\ \delta_k=1;   \\
         &  (x;e_k): x>x_k,\ \mbox{with}\ \delta_k=2;\\
         &  (x,e): e \le x-x_k+e_k, \ \mbox{with}\ \delta_k=3, \\
    \end{array}  
\right. 
\end{equation}
with 	 $\delta_k$ an indicator of termination reason equal to 1, 2 or 3, if a $k$-th employee died, resigned, retired or quitted an employer for any other reason.

It is obvious that the examples of truncated and censored insurance data are not limited to the ones discussed above and they can be countless. It is more difficult to give an example of complete data. We, however, will not do this and move on to the development of methods of multidimensional statistical analysis of truncated and censored data.

\section{Formulating the problem}
\label{sec:3}

Section 2 contains a number of specific examples of truncated-censored data used in the insurance industry, that provide some insight into the process of data generation. However, further development requires a formal description of this process, which includes two separate components --- data truncation and data censoring. To simplify, we will describe them separately.

Let us consider  $X_k= (x_k^1,\ldots,x_k^M )$, $k \in \mathcal K$ a set of mutually independent  $M$ – variate random vectors defined on a common probability space $(\mathcal{X}, {\mathfrak{B}}_{\mathcal{X}}, \textbf{P})$,   where $\mathcal{X}$ is the space of the of  $X$ vector values, $\textbf{P}$ is the probability measure, ${\mathfrak{B}}_{\mathcal{X}}$ – the Borel $\sigma$-algebra of subsets  $\mathcal{X}$, and $\mathcal{K}$ – the index set (finite or infinite). 

It should be noted that the truncation process does not concern the vector  $X$, but the space of its values $\mathcal{X}$. 
Let us consider a random vector  $Y \in \mathcal{Y}$ independent of $X$ and assume that a measurable map $g: \mathcal{Y} \to {\mathfrak {B}}_{\mathcal{X}}$ is given, which assigns a measurable set  $T_k \in {\mathfrak{B}}_{\mathcal{X}}$   to each vector $Y_k \in \mathcal{Y}$, where  $\mathcal{Y}$ is the space of values of  $Y$. 
Let us say that in the process of statistical data generation, the space  $ \mathcal{X}$ is truncated by a random set  $T_k \in {\mathfrak {B}}_{\mathcal{X}}$, $k \in \mathcal{K}$ to the subspace  $\mathcal{X} \setminus T_k$, if as a result of truncation  $\mathcal{X}$ only  $X \in \mathcal{X} \setminus T_k$ become observable. 
In other words, the vector  $X$ is observed in a random condition  $X \in \mathcal{X} \setminus T_k$ associated with its  $k$-th realization   $X_k$, $k \in \mathcal{K}$. The set  $T_k$ will be further called the truncation set or the truncating set.  

Truncated data on the vector  $X$ are generated as follows. First, the vector  $Y_k \in \mathcal{Y}$ is implemented, which by mapping $g$ defines the truncating set  $T_k$, which in turn truncates the original space  $\mathcal{X}$ to the subspace $\mathcal{X} \setminus T_k$. Then the vector  $X_k \in \mathcal{X} \setminus T_k$ is observed, but there is no information about the values and number of realizations of the vector $X$ on the truncating set.  

Thus, the value of the vector $X_k$ and the truncating set  $T_k$ contain all available information about the truncated observation, while the family 
\begin{equation}
\label{equ:7}
\{X_k,T_k \},\ k \in \mathcal{K}  
\end{equation}
forms a truncated sample from the distribution  $\textbf{P}$ provided that  $X \in {\bigcup}_{k=1} ^{\infty}(\mathcal{X}\setminus T_k ) $. Further we assume that ${\bigcup}_{k=1}^\infty(\mathcal{X}\setminus T_k ) =\mathcal{X}$, although this is but a detail.

Unlike truncation the process of censoring is connected only with the vector  $X$. The model for such an experiment can be determined by a measurable map 
\begin{equation}
\label{equ:8}
G: \mathcal{X} \to {\mathfrak {B}}_{\mathcal{X}}, 
\end{equation}
which assigns a certain measurable set   $C_k \in {\mathfrak {B}}_{\mathcal{X}}$  to any vector $X_k \in \mathcal{X}$ such that 
\begin{equation}
\label{equ:9}
    \begin{array}{cl}
         & C_k \ni X_k;  \\
         & G^{-1}(C_k )=\{ X:G(X)=C_k \} = C_k, \ k \in \mathcal{K}.\\
    \end{array}  
\end{equation}
Here we mean that in the process of censoring it is not a specific value of the vector  $X_k, \ k \in \mathcal{K}$ that is observed, but only its image $C_k \in {\mathfrak {B}}_{\mathcal{X}}$ shown in the mapping  $G$. In this case,  $C_k$ stands for a set of possible values of the vector  $X_k$ in the experiment  $G$. 

It follows from the condition (\ref{equ:9}) that the family  ${\{ C_k \} }_{k \in \mathcal{K}}$ forms a space partition $\mathcal{X}$ and induces  on $\mathcal{X}$ a  $\sigma$-algebra ${\mathfrak {B}}_{\mathcal{X}}^{G} \subset {\mathfrak {B}}_{\mathcal{X}}$. 

Using the random variable  $X$ and the system of events  ${\{ C_k \} }_{k \in \mathcal{K}}$,  we construct a new random variable $X^G$ equal to the conditional expected value   $X$ with respect to  ${\mathfrak {B}}_{\mathcal{X}}^{G}$, that is  $X^G=\textbf{E}[X|{\mathfrak {B}}_{\mathcal{X}}^{G}]$. In other words, the value  $X^G$ is obtained from  $X$ by means of “local” conditional averaging, that is, averaging those  $X$, that fall into the set  $C_k$. This means that  $X^G$ is a rough version of  $X$, just as  ${\mathfrak {B}}_{\mathcal{X}}^{G}$ with atoms  $C_k$ is an enlarged version of  $\sigma$-algebra  ${\mathfrak {B}}_{\mathcal{X}}$.

The reason we turned to the conditional expectation  $X^G$ is that the experiment  $G$ actually provides  information about this very value. For the same reason, statistical conclusions about the random value $X$ can be made only indirectly, through the value  $X^G$.	

The grounds for using  $X^G$ as an approximation of $X$ are ensured by the equality  $\textbf{E}[X|B]=\textbf{E}[X^G|B]$, $B \in {\mathfrak {B}}_{\mathcal{X}}^{G}$, whence it follows that the value X is determined with the “accuracy” reduced to the values on the sets $B \in {\mathfrak {B}}_{\mathcal{X}}^{G}$. In particular, the equality is fair
$$
\textbf{P}(X \in B)=\textbf{P}(X^G \in B),  \ B \in {\mathfrak {B}}_{\mathcal{X}}^{G}.
$$
For short, we will often omit the superscript in  $X^G$ and ${\mathfrak {B}}_{\mathcal{X}}^{G}$.

Now applying mapping  (\ref{equ:8}) to the vector $X_k, \ k \in \mathcal{K}$ in (\ref{equ:7})  we get a truncated-censored sample
\begin{equation}
\label{equ:10}
  \{ C_k,T_k \}, \ k \in \mathcal{K}, 
\end{equation}
where  $C_k$ end $T_k$  censoring and truncating sets respectively for the $k$-th   realization on an observed random variable  $X$. Hereafter, no distinction is made between complete and censored observations, and truncated and non-truncated observations, because the complete observation can be regarded as censored with  $C_k=\{ X_k \}$, while the non-truncated as truncated with  $T_k= \emptyset$. 

The objective  is to draw conclusions on unconditional distribution   $\textbf{P}$ of the initial vector  $X$ based on the data of type  (\ref{equ:10}).

\section{Quasi-empirical distribution}
\label{sec:4}

Assume that  $\{ X_k \}, \ k=1, \ldots,n$ is a random sample from the distribution  $\textbf{P}$. Let us consider the empirical distribution  ${\textbf{P}}_n$ on $(\mathcal{X},{\mathfrak {B}}_{\mathcal{X}})$ concentrated at the points  $X_1,\ldots,X_n$,  for which the probability value $X_k$ is assumed to be  $1 / n$. Then for any  $B \in {\mathfrak {B}}_{\mathcal{X}}$ by definition
\begin{equation}
\label{equ:11}
\textbf{P}_n (B)=\frac{1}{n} \sum_{k=1}^n \textbf{I}_{X_k}(B),
\end{equation}
where  $\textbf{I}_{X_k}(B)$ is an indicator equal to 1 if $X \in B$ or 0 if $X \notin B$. It is known that for   
\begin{equation}
\label{equ:12}
\textbf{P}_n (B) \to \textbf{P}(B) \ \ a.s., \ \ as \ \ n \to \infty,
\end{equation}                             
that is, the sequence of empirical distributions  $\textbf{P}_n (B)$ is getting infinitely close to the initial distribution  $\textbf{P}(B)$.

Let us consider the case of censored data. By applying mapping $G$ to $\{ X_k \},k=1,\ldots,n$,  we obtain a censored sample
\begin{equation}
\label{equ:13}
\{C_k \} , \ k=1,\ldots,n.
\end{equation}  
In the future, we will use the shorthand notation $G(X_n )$. Given that  $C_k, \ k=1, \ldots ,n$ shape the space partition $\mathcal{X}$ and the probability value  $X_k$, and, consequently, the probability of its image $G(X_k )=C_k$ is $1 / n$, then using the law of total probability we obtain
\begin{equation}
\label{equ:14}
\textbf{P}_n (B)=\frac{1}{n} \sum_{k=1}^n \textbf{P}_n (B|C_k), \ B \in {\mathfrak{B}}_{\mathcal{X}}.
\end{equation}
In the case of censored data, the probability  $\textbf{P}_n (B|C_k)$ is determined only for   $B \in {\mathfrak {B}}_{\mathcal{X}}^{G}$, so the formula (\ref{equ:14}) matches the expression  (\ref{equ:11}) for the empirical distribution concentrated on the disjoint sets  $C_1, \ldots,C_n$. This makes the estimate  (\ref{equ:14}) trivial. 

A meaningful problem arises if we consider the projection of the vector   $X$ on the subspace $\mathcal{X}$ generated, for example, by its first  $m<M$ components. The fact is that the actuary, as a rule, is interested in certain  $m$ components of the vector  $X$, while other  $M-m$ components are hindering (censoring). This is clearly illustrated by the above discussed examples. Thus, in example 1, the life expectancy of the insured $x$ is of practical interest, while the termination age of the observation  $\tau$ is an hindering parameter. In addition, all the examples actually considered the projection of the vector   $X$ on the subspace  $\mathcal{X}$ generated by one (examples 1 and 2) or two components (examples 3 to 5) the application requires.

Where it is necessary to explicitly specify the dimensionality of the random variable space $\mathcal{X}$ we will use the superscript. So denoting  ${\mathcal{X}}^{M-m}$ the subspace generated by  $M-m$ components and identifying a pair of vectors  $(x_1,\ldots,x_m,0,\ldots,0)$; $(0,\ldots,0$ $,x_{(m+1)},\ldots,x_M )$ with the vector  $(x_1,\ldots,x_M )$, we have
$$
{\mathcal{X}}^M={\mathcal{X}}^m \times {\mathcal{X}}^{M-m}.
$$
Let $\textbf{P}_n (B)$ be a given probability measure for  ${\mathcal{X}}^M$, as before. Let us call the  ${\mathfrak {B}}_{\mathcal{X}}^m$ – Borel $\sigma$-algebra of all subsets  $B \subset {\mathcal{X}}^m$, for which $B \times {\mathcal{X}}^{M-m} \in \mathfrak{B}_{\mathcal{X}}^G$. 

By assumption, ${\mathfrak {B}}_{\mathcal{X}}^m$ contains all the Borel subsets of ${\mathcal{X}}^m$. Then for any  $B \in {\mathfrak {B}}_{\mathcal{X}}^m$  the projection of measure $\textbf{P}_n$ on ${\mathcal{X}}^m$ is
$$
\textbf{P}_n (B)=\textbf{P}_n (B \times {\mathcal{X}}^{M-m} ).
$$
Hence, the projection  (\ref{equ:14}) on  ${\mathcal{X}}^m$ is a functional
\begin{equation}
\label{equ:15}
\textbf{P}_n (B)=\frac{1}{n} \sum_{k=1}^n \textbf{P} (B|C_k), \ B \in {\mathfrak {B}}_{\mathcal{X}}^m, \ \textbf{P} \in \mathcal{P},
\end{equation}
where $\mathcal{P}$ is the class of all distributions representing the projection of measure   $\textbf{P}$ from  ${\mathcal{X}}^M$ on  ${\mathcal{X}}^m$, such that    $\textbf{P}(C_k) = {1} /{n} $,  $C_k \in \mathfrak{B}_{\mathcal{X}}^G$. 

Note that the function  (\ref{equ:14}) is projected into the functional (\ref{equ:15}). This is because the censored sample (\ref{equ:13}) does not regulate the distribution of measure  $\textbf{P}_n$ on $C_k \in \mathfrak{B}_{\mathcal{X}}^G$ and the projections of censoring sets $C_k$ on ${\mathcal{X}}^m$ intersect. Indeed, the calculation of probabilities  $\textbf{P}(B|C_k )$, $B \in \mathfrak{B}_{\mathcal{X}}^m$ assumes the distribution  $\textbf{P}$ on the sets  $C_k \in \mathfrak{B}_{\mathcal{X}}$ as known. Class $\mathcal{P}$ contains infinitely many distributions $\textbf{P}$, which implies the ambiguity of the projection of  $\textbf{P}_n(B)$ on ${\mathcal{X}}^m$. 
According to the paper   [3] the functional solution  (\ref{equ:15}), that is
\begin{equation}
\label{equ:16}
{\textbf{P}}_n^* (B)=\frac{1}{n} \sum_{k=1}^n {\textbf{P}}_n^* (B|C_k), \ B \in {\mathfrak {B}}_{\mathcal{X}}^m, \ {\textbf{P}}_n^* \in \mathcal{P},
\end{equation}
is called quasi-empirical distribution. Obviously, such a solution exists. For example, if we assume that on  $C_k \in \mathfrak{B}_{\mathcal{X}}^G$,  the measure  $\textbf{P}_n (C_k)= {1} /{n} $ is concentrated at the point  $X_k \in C_k$, then (\ref{equ:14}) transforms into the empirical distribution (\ref{equ:11}), and the solution of the functional (\ref{equ:15}) will be usual empirical distribution on ${\mathcal{X}}^m$. 

The following treatment concerns only the vector  $X^m=(x_1,\ldots,$ $x_m,$ $0,\ldots,0)$, and therefore we will omit the index $m$ used to denote the dimensionality of the vector  ${X}^m$, the subspace  ${\mathcal{X}}^m$, and $\sigma$-algebra ${\mathfrak {B}}_{\mathcal{X}}^m$,  if no misunderstanding follows.

Let us consider the complete truncated sample (\ref{equ:7}) from the distribution  $\textbf{P}$. Assume that  $T_k \ne \mathcal{X}$ for any  $k \in \mathcal{K}$, then
$$
\textbf{P}(B|\mathcal{X}\setminus T_k)=\frac{\textbf{P}(B \cap (\mathcal{X}\setminus T_k))}{\textbf{P}(\mathcal{X}\setminus T_k)})=\frac{\textbf{P}(B )- \textbf{P}(B \cap T_k)}{1-\textbf{P}(T_k)}.
$$
Whence it follows that
\begin{equation}
\label{equ:17}
\frac{\textbf{P}(B)}{1-\textbf{P}(\ T_k)}=\textbf{P}(B|\mathcal{X}\setminus T_k)+\frac{\textbf{P}(B \cap T_k)}{1-\textbf{P}(T_k)}. 
\end{equation}
Having summed up the left and right parts of this expression across all $k=1,\ldots,n$, we obtain
\begin{equation}
\label{equ:18}
\textbf{P}(B)=\frac{1}{N} \sum_{k=1}^n \left(\textbf{P}(B|\mathcal{X}\setminus T_k)+\frac{\textbf{P}(B \cap T_k)}{1-\textbf{P}(T_k)}\right),\ B \in {\mathfrak {B}}_{\mathcal{X}}^m, \ \textbf{P} \in \mathcal{P} , 
\end{equation}
where $N$ is the sample size adjusted for truncation, equal to
$$
N=\sum_{k=1}^n \frac{1}{1-\textbf{P}(T_k)}.      
$$    
Now assume that observation is censored using the censoring set  $C_k$. Given that  $C_k \cap T_k=\emptyset$, then  $\textbf{P}(B|C_k; \mathcal{X}\setminus T_k)$= $\textbf{P}(B|C_k)$, \ $B \in {\mathfrak {B}}_{\mathcal{X}}^m$, \ $\textbf{P} \in \mathcal{P}$, therefore
\begin{equation}
\label{equ:19}
\textbf{P}_n(B)=\frac{1}{N} \sum_{k=1}^n \left(\textbf{P}(B|C_k)+\frac{\textbf{P}(B \cap T_k)}{1-\textbf{P}(T_k)}\right),\ B \in {\mathfrak {B}}_{\mathcal{X}}^m, \ \textbf{P} \in \mathcal{P} . 
\end{equation}

The solution of this functional 
\begin{equation}
\label{equ:20}
\textbf{P}_n^*(B)=\frac{1}{N} \sum_{k=1}^n \left(\textbf{P}_n^*(B|C_k)+\frac{\textbf{P}_n^*(B \cap T_k)}{1-\textbf{P}_n^*(T_k)}\right),\ B \in {\mathfrak {B}}_{\mathcal{X}}^m, \ \textbf{P}_n^* \in \mathcal{P} . 
\end{equation}
we will call quasi-empirical distribution (or qED),  since it actually generalizes a similar expression of quasi-empirical distribution for censored data  (\ref{equ:16}). Indeed, if in  (\ref{equ:10}) we assume that  $T_k \equiv \emptyset, \ k=1,\ldots,n$, then ${\textbf{P}_n^*(B \cap T_k)}/$ ${(1-\textbf{P}_n^*(T_k))}=0$ and $N=n$. 

The expression  (\ref{equ:20}) allows the interpretation as follows. Assume that the collection of complete data on the vector  $X$ from the distribution  $\textbf{P}(B)$ terminates immediately after the event  $X_k \in \mathcal{X} \setminus T_k$ occurs for the first time. 
Then on the set  $T_k$,  ${\textbf{P}(T_k )}/{(1-\textbf{P}(T_k))}$ realizations of this vector  will be observed. 
Note that this corresponds to the collection of truncated data with the only difference, that in the latter case there is no information about the values and number of realizations of the vector  $X$ on the truncation set  $T_k$. However, the lack of information does not affect the probabilistic characteristics of the vector $X$. 
Therefore, to convert one truncated observation  $X_k \in \mathcal{X} \setminus T_k$ from the distribution $\textbf{P}(B \mid {\mathcal{X} \setminus T_k })$ to a non-truncated observation from the distribution  $\textbf{P}(B)$, it is necessary to add ${\textbf{P}(T_k )}/{(1-\textbf{P}(T_k))}$ unobserved realizations to one observed realization $X_k \in \mathcal{X} \setminus T_k$, that is, each truncated observation is replaced by an equivalent number 
%$\mathcal{X} \in T_k$
\begin{equation}
\label{equ:21}
1+\frac{\textbf{P}(T_k)}{1-\textbf{P}(T_k)}=\frac{1}{1-\textbf{P}(T_k)}, 
\end{equation}
of non-truncated observations from distribution $\textbf{P}(B)$. As a result, a non-truncated sample is formed from the distribution  $\textbf{P}(B)$, on which the estimate  $\textbf{P}_n (B)$ is built, similar to the usual empirical distribution.

Let us explain the meaning of the second summand in expression  \ (\ref{equ:20}): the numerator  $\textbf{P}_n^* (B \cap T_k)$ “allocates” to the set $B$ a part of the probability measure concentrated on the truncation set  $T_k, \ k=1,\ldots,n$, and the denominator  $1- \textbf{P}_n^* (T_k)$ normalizes it, meeting the condition  (\ref{equ:21}). 

Note that if the equality  (\ref{equ:17}) is multiplied  $1- \textbf{P} (T_k)$ and the previous arguments are repeated, we obtain an equivalent estimate (in a certain sense) of the qED:
\begin{eqnarray}
\label{equ:22}
\textbf{P}_n^*(B)=\frac{1}{n} \sum_{k=1}^n \left(\textbf{P}_n^*(B|C_k) {(1-\textbf{P}_n^*(T_k))}+ \right.
\nonumber\\ \left. {\textbf{P}_n^*(B \cap T_k)}\right), 
\ B \in {\mathfrak {B}}_{\mathcal{X}}^m, \ \textbf{P}_n^* \in \mathcal{P} . 
\end{eqnarray}

Further, let  $\mathcal{X}$ include sets  $B_t=\{ X \in \mathcal{X}: x_k < t_k, \ k \in \mathcal{K} \}$, where $t \in \mathcal{X}$, then the function  $\textbf{F}_n^* (t)=\textbf{P}_n^* (B_t )$ is a distribution function in the popular sense, that is, the unique, unambiguous, real, and non-negative function of the point  $t \in \mathcal{X}$. The function  $\textbf{F}_n^* (t)$ will be called a quasi-empirical distribution function.

In this paper, we will not touch upon such a complex and important issue as the dependence of the individual components of the vector  $X_k=(x_k^1,\ldots,x_k^m )$. For the sake of simplicity, we assume that all the components involved in the process of censoring and truncating observations, but not explicitly included in the list of components of the estimated distribution function, are independent of the latter.

To construct an estimate of the distribution function  (\ref{equ:20}) or (\ref{equ:22}), one can use an iterative procedure of EM-algorithm (e.g., Dempster et al.~\cite{Dempster1977}), which includes the following steps.

	1. 	Set the initial approximation of the estimate  $\textbf{P}_n^{(0)}(B), \ B \in \mathcal{X}$, for example, uniform.
	
	2. 	Calculate the sample size  adjusted for truncation by formula
$$
\qquad N^{(0)}= \sum_{k=1}^n \frac{1}{1-\textbf{P}_n^{(0)}(T_k)}, 
$$

    3. 	Calculate the new value   $\textbf{P}_n^{(1)}(B)$ using the formula
$$
\qquad \textbf{P}_n^{(1)} (B)=\frac {1}{N^{(0)}} \sum_{k=1}^n \left(\textbf{P}_n^{(0)} (B \mid C_k) + \frac {\textbf{P}^{(0)}(B \cap T_k)}{ (1-\textbf{P}_n^{(0)}(T_k))}\right)
$$
\qquad \quad or
$$
\qquad \textbf{P}_n^{(1)} (B)=\frac{1}{n} \sum_{k=1}^n \left(\textbf{P}_n^{(0)} (B \mid C_k)  \cdot (1-\textbf{P}_n^{(0)}(T_k)) +  \textbf{P}_n^{(0)}(B \cap T_k)\right).
$$
% \qquad \quad \looseness=-1 если квазиэмпирическое распределение определено в форме  (\ref{equ:22}). 

	4. Return to step   (2), replacing   $\textbf{P}_n^{(0)} (\cdot)$ with $\textbf{P}_n^{(1)} (\cdot)$, etc.
	
	5. Calculations are complete with the set accuracy achieved.

\section{Estimation of univariate distribution function (general case)}
\label{sec:5}

Let us consider the truncated and censored sample (\ref{equ:10}) from the distribution $\textbf{F}(x)$ of a univariate random variable $x$. The censoring and truncating sets are respectively equal to
\begin{equation}
\label{equ:23}
    \begin{array}{cl}
         & C_k =(c_k^1,c_k^2 );  \\
         & T_k=(-\infty,t_k^1 )\cup (t_k^2,\infty),\\
    \end{array}  
\end{equation}
and their boundaries are connected by $t_k^1 \le c_k^1<c_k^2 \le t_k^2, \ k=1,\ldots,n$. Note that special cases of this type of data: left-truncated and right-censored (examples 1 and 2) with $T_k=(-\infty,t_k^1)$ and $C_k=(c_k^1,\infty)$ were considered previously.

It's obvious that the range of univariate truncated and censored data generated by the sets (\ref{equ:23}) is much wider than those previously considered. It is possible to distinguish 12 different types of truncated and censored observations listed in Table 1. Note that not all combinations of censoring and truncation are possible. In the table invalid combinations are marked by $(-)$, they include right (left)-censored and right (left)-truncated or doubly truncated observations.

\begin{table}
\begin{center}
\caption{\label{tab:1} Various types of truncated and censored observations}
\begin{tabular}{lcccc}
\noalign{\smallskip}\hline\noalign{\smallskip}

& \multicolumn{4}{c}{Truncation} \\
\cline{2-5}
\raisebox{1.9ex}[0.4cm][0.1cm]{Censoring~~~~~~~}
     &  ~~~~~~~No~~~~~~~ & ~~~~~~Right~~~~~ & ~~~~~~Left~~~~~~ & ~~~~~Double~~~~~\\
\noalign{\smallskip}\hline\noalign{\smallskip}
No (complete)    & + & +  & +  & +  \\
Right            & + & -- & +  & -- \\
Left             & + & +  & -- & -- \\
Interval         & + & +  & +  & +  \\
\noalign{\smallskip}\hline
\end{tabular}
\end{center}
\end{table}

The qED function $\textbf{F}_n^*(x)$, with all sample elements  double-truncated and interval-censored, i.e., with all $C_k$ and $T_k, \ k=1, \ldots ,n$ looking as (\ref{equ:23}), is a solution of the functional\footnote{Further we will consider quasi-empirical distribution (\ref{equ:20}).}
\begin{eqnarray}
\label{equ:24}
%\begin{split}
\textbf{F}_n(x)=\frac{1}{N} \sum_{k=1}^n \left(\frac{\textbf{F}\left(\min(c_k^2,x)\right)-\textbf{F}\left(\min(c_k^1,x)\right)}{\textbf{F}(c_k^2)-\textbf{F}(c_k^1)}\right. +\nonumber\\  
 \left. \frac{\textbf{F}\left(\max(x,t_k^2)\right)-\textbf{F}(t_k^2)+\textbf{F}\left(\min(x,t_k^1)\right)}{\textbf{F}(t_k^2)-\textbf{F}(t_k^1)}\right)
 %\end{split}
\end{eqnarray}
with	 $N$ an adjusted sample size equal to 
\begin{equation}
N= \sum_{k=1}^n \frac{1}{\textbf{F}(t_k^2 )-\textbf{F}(t_k^1 )}. 
\end{equation}
Note that in the special cases of censoring and truncation (see Table \ref{tab:1}) certain boundaries of sets (\ref{equ:23}) are changed to their limits, which allows to simplify the expressions for summands in (\ref{equ:24}). So, with $k$-th observation

--- complete,   we put $c_k^1 \le x_k \le c_k^2$, $c_k^1 \to x_k$ and $c_k^2 \to x_k$, then $C_k=\{x_k \}$ and 
$$
\lim_{c_k^1 \to x_k; \ c_k^2 \to x_k}\left(\frac{\textbf{F}\left(\min(c_k^2,x)\right)-\textbf{F}\left(\min(c_k^1,x)\right)}{\textbf{F}(c_k^2)-\textbf{F}(c_k^1)}\right)=\textbf{I}(x_k \le x);
$$

--- right-censored, then $c_k^2=t_k^2=\infty$ \ and \ $\textbf{F}(c_k^2 )=\textbf{F}(\max(x,t_k^2 ) )= \ \textbf{F}(t_k^2 )=1$;

--- left-censored, then $t_k^1=c_k^1=-\infty$ and $\textbf{F}(\min(c_k^1,x) )=\textbf{F} ( \min(x,t_k^1 ) )=\textbf{F}(t_k^1 )=0$;

--- right nontruncated, then $t_k^2=\infty$ and $\textbf{F}(t_k^2 )=1$;

--- left nontruncated, then  $t_k^1=-\infty$ and $\textbf{F}(t_k^1 )=0$.

Formulae for the summands in (\ref{equ:24}) for various combinations of truncation and censoring of observations, can be found in Table  \ref{tab:2}.

\begin{table}
%[ph]
  \caption{\label{tab:2} Summands of quasi-empirical distribution depending on the observation type}
    \centering
  \begin{tabular}{l c}

\hline\noalign{\smallskip}
\multicolumn {1}{c}{Observation type} & \multicolumn {1}{c}{Formula} \\
\noalign{\smallskip}\hline\noalign{\smallskip}
    \multicolumn {1}{l}{Complete and } &  \multirow{2}{*}{$\textbf{I}(x_k \le x)$}\\
    \multicolumn {1}{l}{Nontruncated }\\
\noalign{\smallskip}\hline\noalign{\smallskip}
    \multicolumn {1}{l}{Right-Censored and } &  \multirow{2}{*}{$\frac{\textbf{F}(x)-\textbf{F}\left(\min(c_k^1,x)\right)}{1-\textbf{F}(c_k^1)}$}\\
    \multicolumn {1}{l}{Nontruncated }\\
\noalign{\smallskip}\hline\noalign{\smallskip}
    \multicolumn {1}{l}{Left-Censored and } &  \multirow{2}{*}{$\frac{\textbf{F}\left(\min(c_k^2,x)\right)}{\textbf{F}(c_k^2)}$}\\
    \multicolumn {1}{l}{Nontruncated }\\
\noalign{\smallskip}\hline\noalign{\smallskip}
    \multicolumn {1}{l}{Interval-censored and } &  \multirow{2}{*}{$\frac{\textbf{F}\left(\min(c_k^2,x)\right)-\textbf{F}\left(\min(c_k^1,x)\right)}{\textbf{F}(c_k^2)-\textbf{F}(c_k^1)}$}\\
    \multicolumn {1}{l}{Nontruncated }\\
\noalign{\smallskip}\hline\noalign{\smallskip}
    \multicolumn {1}{l}{Complete and } &  \multirow{2}{*}{$\textbf{I}(x_k \le x)+\frac{\textbf{F}\left(\min(x,t_k^1)\right)}{1-\textbf{F}(t_k^1)}$}\\
    \multicolumn {1}{l}{Left-Truncated }\\
\noalign{\smallskip}\hline\noalign{\smallskip}
    \multicolumn {1}{l}{Right-Censored and } &  \multirow{2}{*}{$\frac{\textbf{F}(x)-\textbf{F}\left(\min(c_k^1,x)\right)}{1-\textbf{F}(c_k^1)}+\frac{\textbf{F}\left(\min(x,t_k^1)\right)}{1-\textbf{F}(t_k^1)}$}\\
    \multicolumn {1}{l}{Left-Truncated }\\
\noalign{\smallskip}\hline\noalign{\smallskip}
    \multicolumn {1}{l}{Left-Censored and } &  \multirow{2}{*}{---}\\
    \multicolumn {1}{l}{Left-Truncated }\\
\noalign{\smallskip}\hline\noalign{\smallskip}
    \multicolumn {1}{l}{Interval-Censored and } &  \multirow{2}{*}{$\frac{\textbf{F}\left(\min(c_k^2,x)\right)-\textbf{F}\left(\min(c_k^1,x)\right)}{\textbf{F}(c_k^2)-\textbf{F}(c_k^1)}+\frac{\textbf{F}\left(\min(x,t_k^1)\right)}{1-\textbf{F}(t_k^1)}$}\\
    \multicolumn {1}{l}{Left-Truncated }\\
\noalign{\smallskip}\hline\noalign{\smallskip}
    \multicolumn {1}{l}{Complete and } &  \multirow{2}{*}{$\textbf{I}(x_k \le x)+\frac{\textbf{F}\left(\max(x,t_k^2)\right)-\textbf{F}(t_k^2)}{\textbf{F}(t_k^2)}$}\\
    \multicolumn {1}{l}{Right-Truncated }\\
\noalign{\smallskip}\hline\noalign{\smallskip}
    \multicolumn {1}{l}{Right-Censored and } &  \multirow{2}{*}{---}\\
    \multicolumn {1}{l}{Right-Truncated }\\
\noalign{\smallskip}\hline\noalign{\smallskip}
    \multicolumn {1}{l}{Left-Censored and } &  \multirow{2}{*}{$\frac{\textbf{F}\left(\min(c_k^2,x)\right)}{\textbf{F}(c_k^2)}+\frac{\textbf{F}\left(\max(x,t_k^2)\right)-\textbf{F}(t_k^2)}{\textbf{F}(t_k^2)}$}\\
    \multicolumn {1}{l}{Right-Truncated }\\
\noalign{\smallskip}\hline\noalign{\smallskip}
    \multicolumn {1}{l}{Interval-Censored and } &  \multirow{2}{*}{$\frac{\textbf{F}\left(\min(c_k^2,x)\right)-\textbf{F}\left(\min(c_k^1,x)\right)}{\textbf{F}(c_k^2)-\textbf{F}(c_k^1)}+\frac{\textbf{F}\left(\max(x,t_k^2)\right)-\textbf{F}(t_k^2)}{\textbf{F}(t_k^2)}$}\\
    \multicolumn {1}{l}{Right-Truncated }\\
\noalign{\smallskip}\hline\noalign{\smallskip}
    \multicolumn {1}{l}{Complete and } &  \multirow{2}{*}{$\textbf{I}(x_k \le x)+\frac{\textbf{F}\left(\max(x,t_k^2)\right)-\textbf{F}(t_k^2)+\textbf{F}\left(\min(x,t_k^1)\right)}{\textbf{F}(t_k^2)-\textbf{F}(t_k^1)}$}\\
    \multicolumn {1}{l}{Double-Truncated }\\
\noalign{\smallskip}\hline\noalign{\smallskip}
     \multicolumn {1}{l}{Right-Censored and } &  \multirow{2}{*}{---}\\
     \multicolumn {1}{l}{Double-Truncated }\\
\noalign{\smallskip}\hline\noalign{\smallskip} 
     \multicolumn {1}{l}{Right-Censored and } &  \multirow{2}{*}{---}\\
     \multicolumn {1}{l}{Double-Truncated }\\
\noalign{\smallskip}\hline\noalign{\smallskip}
       \multicolumn {1}{l}{Interval-Censored,  } &  \multirow{2}{*}{$\frac{\textbf{F}\left(\min(c_k^2,x)\right)-\textbf{F}\left(\min(c_k^1,x)\right)}{\textbf{F}(c_k^2)-\textbf{F}(c_k^1)} +\frac{\textbf{F}\left(\max(x,t_k^2)\right)-\textbf{F}(t_k^2)+\textbf{F}\left(\min(x,t_k^1)\right)}{\textbf{F}(t_k^2)-\textbf{F}(t_k^1)} $}\\
      \multicolumn {1}{l}{Double-Truncated }\\
\noalign{\smallskip}\hline\noalign{\smallskip}
  \end{tabular}
\end{table}

\section{Accuracy of estimation of quasi-empirical distribution function 
(Univariate survival data)}
\label{sec:6}

Simulation study was organized to verify the accuracy of estimation of the qED function for small samples. 
This is a rather complex challenge due to a large number of factors affecting the accuracy of estimation of the distribution function based on truncated-censored data. 
In addition to traditional factors such as the probability distributing the original random variables and the sample size, the estimation accuracy  in this case is also affected by the distribution of certain components of the vectors   $X$ and $Y$, as well as their mapping  $G$ and $g$ (see sec.~ \ref{sec:3}). These factors determine the structure of the truncated-censored sample  (\ref{equ:10}), which is characterized by the type of truncating and censoring sets, their number, and relative position in the sample. The structure of such sample can be approximated by the following indicators: 

--- the type of truncating and censoring sets available in the sample (possible variants of truncated-censored observations are given in Table~\ref{tab:1}); 

--- the degree of data truncation and the degree of censoring (the proportion of truncated and censored observations in the sample), differentiated by types of truncation and censoring sets.

Based thereon, the simulation program provided for varying the structure of the truncated-censored sample within a wide range. Samples from the distribution  $\textbf{F}(x)$, containing only certain types of truncated-censored observations presented in Table~\ref{tab:2}, as well as their various combinations, were considered. The degree of censorship and truncation varied from 0 to 0.85, and the sample size varied from 100 to 5000 observations. Random variables from Gamma, Lognormal, Tweedie, and Weibull distributions were studied. 

In addition to examining the influence of these parameters on the accuracy of qED estimator, the simulation program provided for its comparison with the modified Turnbull estimator~\cite{Frydman1994}, and in the case of truncated-left and censored-right samples, with the modified Kaplan-Meier estimator~\cite{Lai1991}.

Estimating accuracy of some  statistics $s$ (for example, quantile or mean) based on truncated-censored data, relative error  $\delta = \hat s / s$ and the corresponding confidence interval $(\delta_{1- {\alpha} / {2} }, \ \delta_{ {\alpha}/ {2} } )$, such that $\textbf{P}(\delta_{1-\alpha /2}, \ \delta_{\alpha / 2} )=\alpha$, where  $\hat s$ estimation of statistic $s$, were used as the criteria. The accuracy of the estimation of cumulative distribution function $\textbf{F}(x)$ is integrally defined by Chebyshev distance
\begin{equation}
\label{equ:25}
\rho=\max_{n} | \textbf{F}_n^* (x)-\textbf{F}(x) |, 
\end{equation}
error of the mean $\textbf{E} [ \textbf{F}_n^* (x) ]$ and two-sided confidence bands, where $\textbf{F}_n^* (x)$ is qED or alternative estimation of cumulative distribution function $\textbf{F}(x)$.

Let us consider several examples illustrating the influence of different parameters on the accuracy of qED estimator. The results of simulation obtained make no claims to completeness, but in our opinion, allow to conclude whether the proposed evaluation of the distribution function based on truncated-censored data is effective. 

\begin{figure}
\center
\includegraphics[scale=0.7]{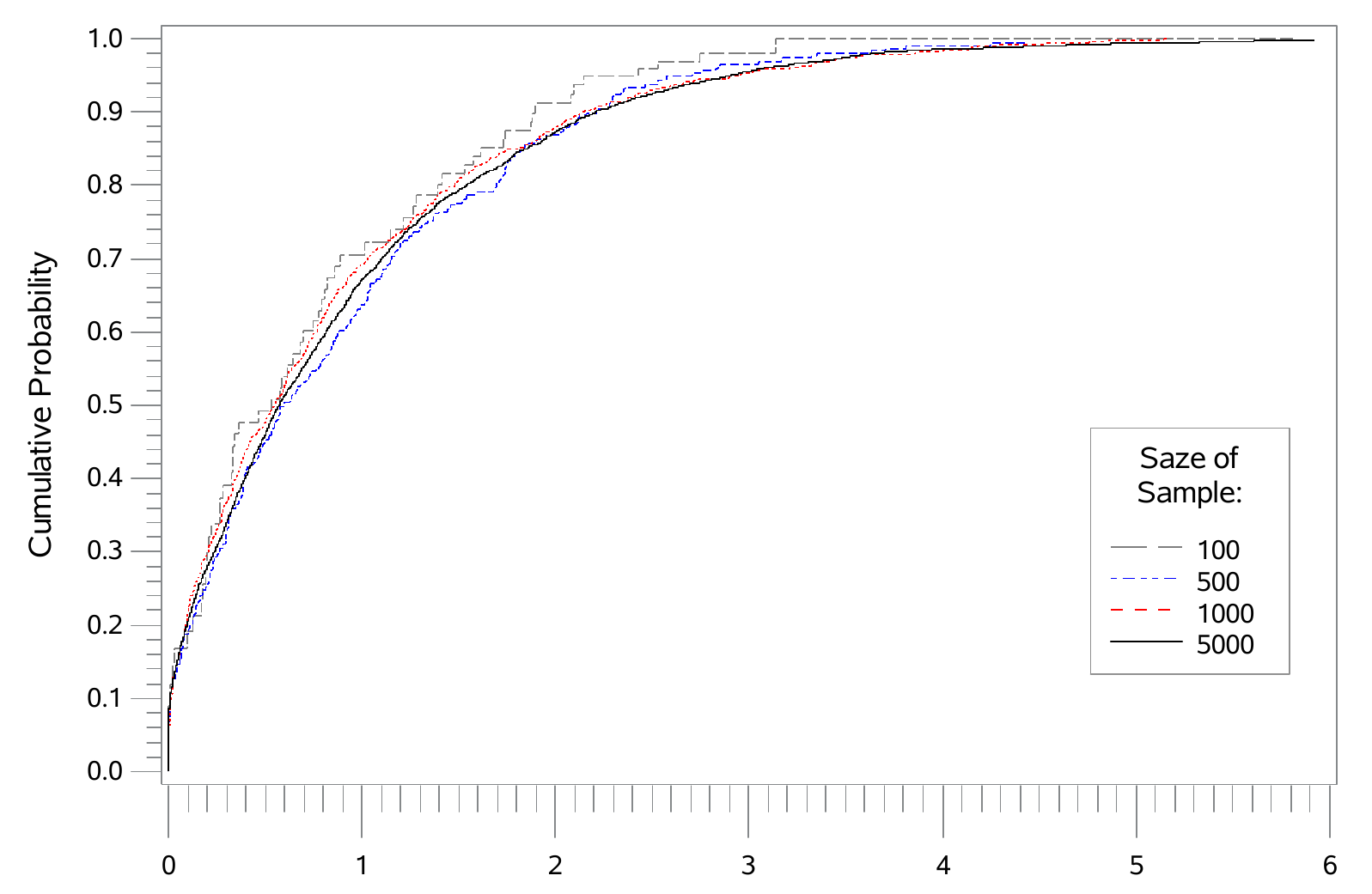}
\caption{Nonparametric estimation of the distribution function for left-truncated and right-censored sample depending on its size}
\label{fig:5}       % Give a unique label
\end{figure}

\paragraph{Example 6.} Fig.~\ref{fig:5} gives the examples of constructing an estimate of the univariate distribution function based on left-truncated and right-censored data from the Tweedie distribution~\cite{Tweedie1984} with an index parameter $p=1.7$, expected value, and dispersion equal to  1. In the process of simulation, the size of the truncated sample $n$ took the values of 100, 500, 1000, and 5000 observations, while other parameters of the model remained unchanged. As a result, the main characteristics of these samples were quite close to each other. Thus, the degree of truncation varied from 0.61 to 0.68, and the degree of censoring varied from 0.36 to 0.41. The size of the non-truncated sample is approximately equal to $N \approx 1.55 \cdot n$.

The results of calculations show that with the growth of the sample size, the estimate of the qED stabilizes and virtually matches the modified Kaplan-Meier estimator. 
For example, at $n \geq  1000$ the Chebyshev distance between the given estimates does not exceed  $3.65 \cdot 10^{-7}$. 

\begin{figure}
\center
  \includegraphics[scale=0.7]{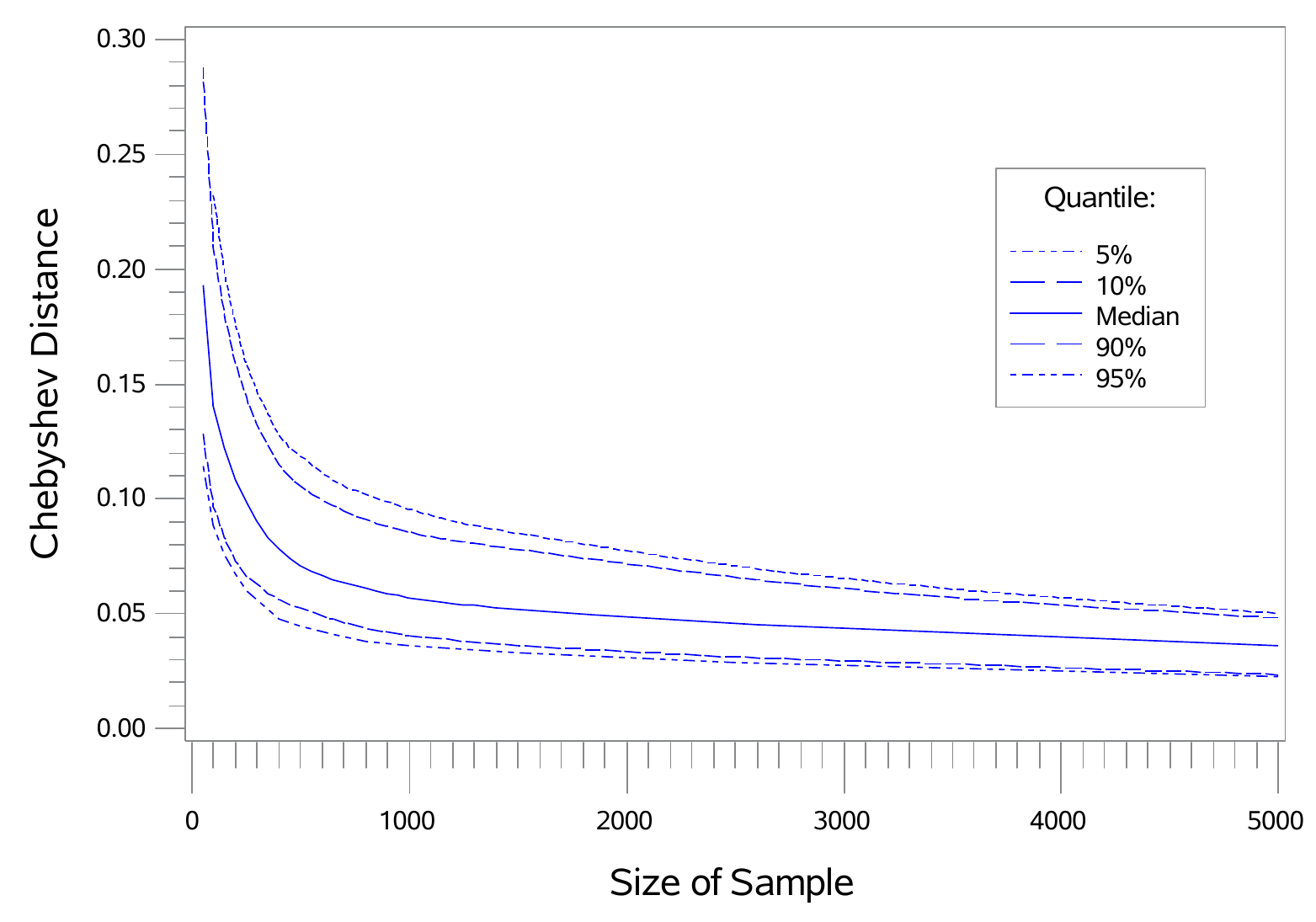}
\caption{The dependence of quantiles of the Chebyshev distance distribution from the size of a truncated-censored sample (from top to bottom: 95\%, 90\%, 50\%, 10\%, and 5\% quantiles)}
\label{fig:6}       % Give a unique label
\end{figure}

\paragraph{Example 7.} The effect of sample size on the estimation accuracy of the qED for truncated and censored sample from the Weibull distribution with the shape parameter  $a=5$ and the scale parameter $\lambda=10$ is illustrated by Fig.~\ref{fig:6}. 
The sample structure is shown in Table~\ref{tab:3}. 
The degree of  observations censoring in the example is 0.8, and the degree of truncation is 0.55 on average. 
The size of truncated and censored sample $n$ varied from 50 to 5000 observations, and the size of the corresponding complete sample is approximately equal to $N \approx 1.98 \cdot n$.

\begin{table}
\begin{center}
\caption{\label{tab:3} Truncated-censored sampling structure}
\begin{tabular}{lcc}
\noalign{\smallskip}\hline\noalign{\smallskip}

& \multicolumn{2}{c}{Percentage share} \\  \cline{2-3}

\raisebox{1.9ex}[0.4cm][0.1cm]{Type of truncated and censored observations~~~}
     &  ~~Example 7~~ & ~~Example 9~~ \\
\noalign{\smallskip}\hline\noalign{\smallskip}
Complete and Nontruncated	             &   0	&  10  \\
Right-Censored and Nontruncated	         &  15	&  10  \\
Left-Censored and Nontruncated           &  15  &  10  \\
Interval-Censored and Nontruncated	     &  15  &	0  \\
Complete and Left-Truncated	             &  10	&  10  \\
Right-Censored and Left-Truncated	     &  10	&  10  \\
Interval-Censored and Left-Truncated     &  10	&  10  \\
Complete and Right-Truncated	         &   5	&  10  \\
Left-Censored and Right-Truncated	     &   5	&  10  \\
Interval-Censored and Right-Truncated    &   5	&   0  \\
Complete and Doubly Truncated	         &   5	&  10  \\
Interval-Censored and Doubly Truncated   &   5	&  10  \\
\noalign{\smallskip}\hline
\end{tabular}
\end{center}
\end{table}

For each value  $n$,  500 truncated and censored samples were simulated to evaluate qED. Then, the Chebyshev distance ($\rho$ ) between this estimate and the true  Weibull distribution was calculated by the formula (\ref{equ:25}). For each examined value $n$, the distribution of distance $\rho$ was constructed and quantiles were estimated. 

The figure shows the dependence of 95\%, 90\%, 50\%, 10\% and 5\% quantiles of distribution of distance  $\rho$ from the sample size.  The median and other quantiles of the specified distribution decrease monotonically with the growth of  $n$. 
Moreover, the rate of decline in the indicators  depends on $n$ --- the lesser is  $n$, the higher is the rate of decline. 
For example, with an increase of  $n$ from 50 to 500, that is, by  450 observations, the median decreases from 0.201 to 0.075, or 2.68 times. 
With an increase of $n$ from 500 to 2500, that is, by 2000 observations, the median value decreases only 1.59 times. 
At the same time, the standard deviation with the growth of $n$ decreases at a higher rate. Thus, at $n$ equal to 50, 500, and 2500, the standard deviation is respectively 0.064, 0.031, and 0.0.12, that is, decreases 2.06 and 2.58 times. 

From a practical point of view, the revealed patterns mean that with the growth of a truncated-censored sample size, the accuracy of estimation of qED uniformly increases across the space of the random variable values and, at the same time, the variance of estimate decreases. 
It is obvious that with the increase of $n$, growth rates of the estimation accuracy of individual quantiles is higher than previously specified for the distribution function as a whole. 
For example, the maximum relative error of the median at $n=50$ is 31.0\%, and at $n=2500$ the error does not exceed 3.3\%. 

Thus, it can be concluded that with a size of truncated and censored sample, commensurate with that of the insurance companies data, the accuracy of qED estimation  becomes acceptable for most applications.

\begin{landscape}
\begin{figure*}
    \begin{center}
    \begin{tabular}{cccc}
    \includegraphics[width=0.37\textwidth]{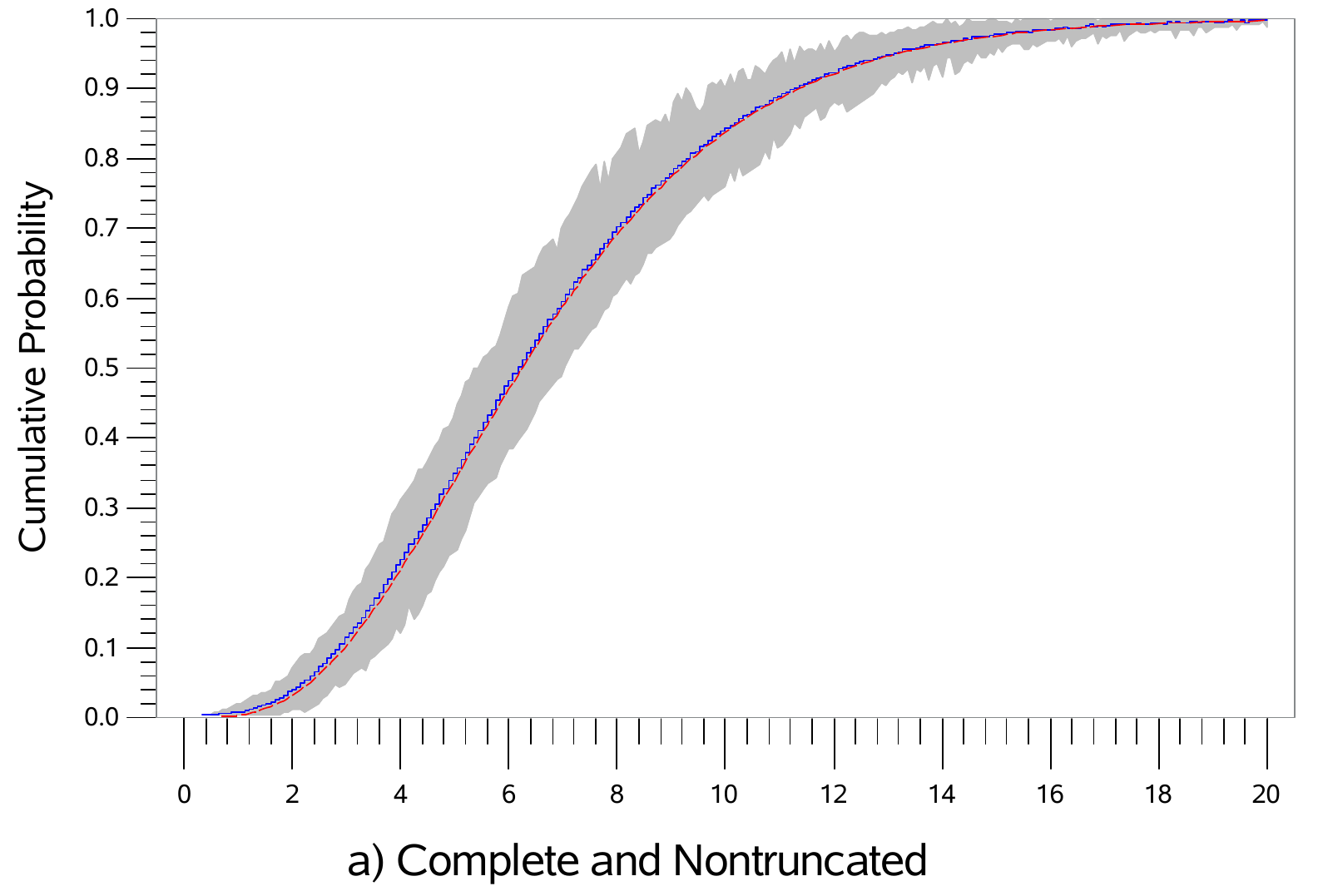} &
    \includegraphics[width=0.37\textwidth]{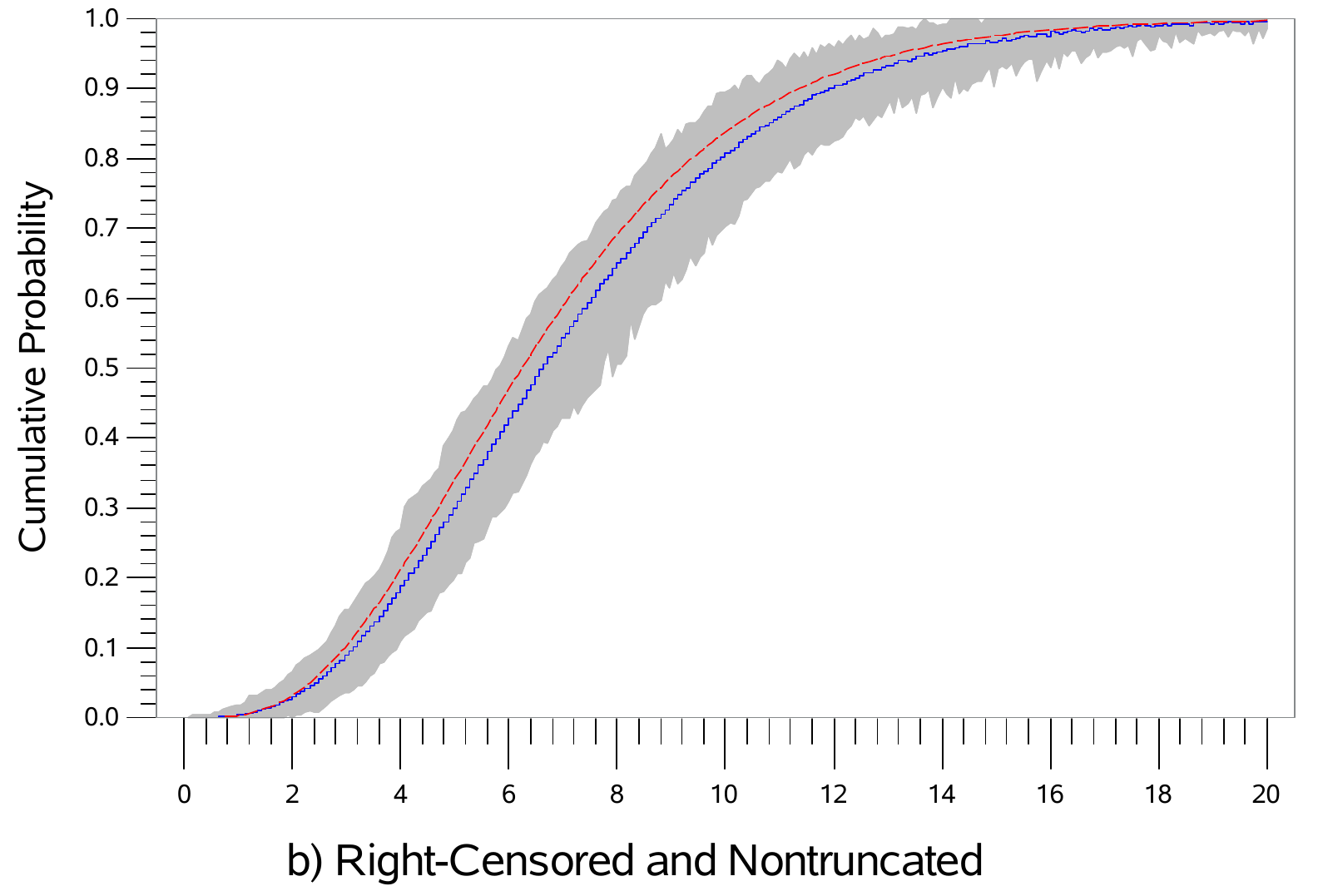} &
    \includegraphics[width=0.37\textwidth]{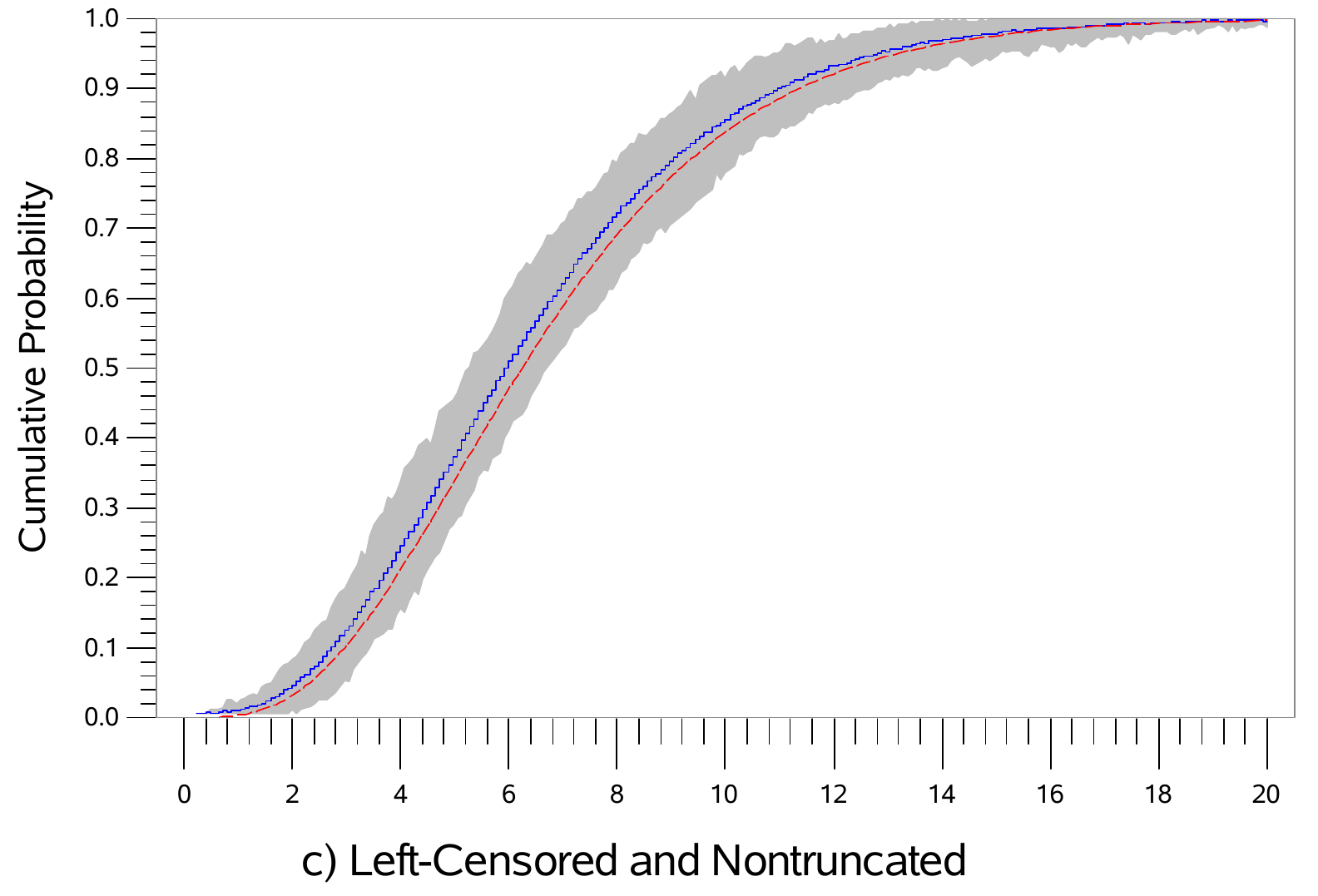} &
    \includegraphics[width=0.37\textwidth]{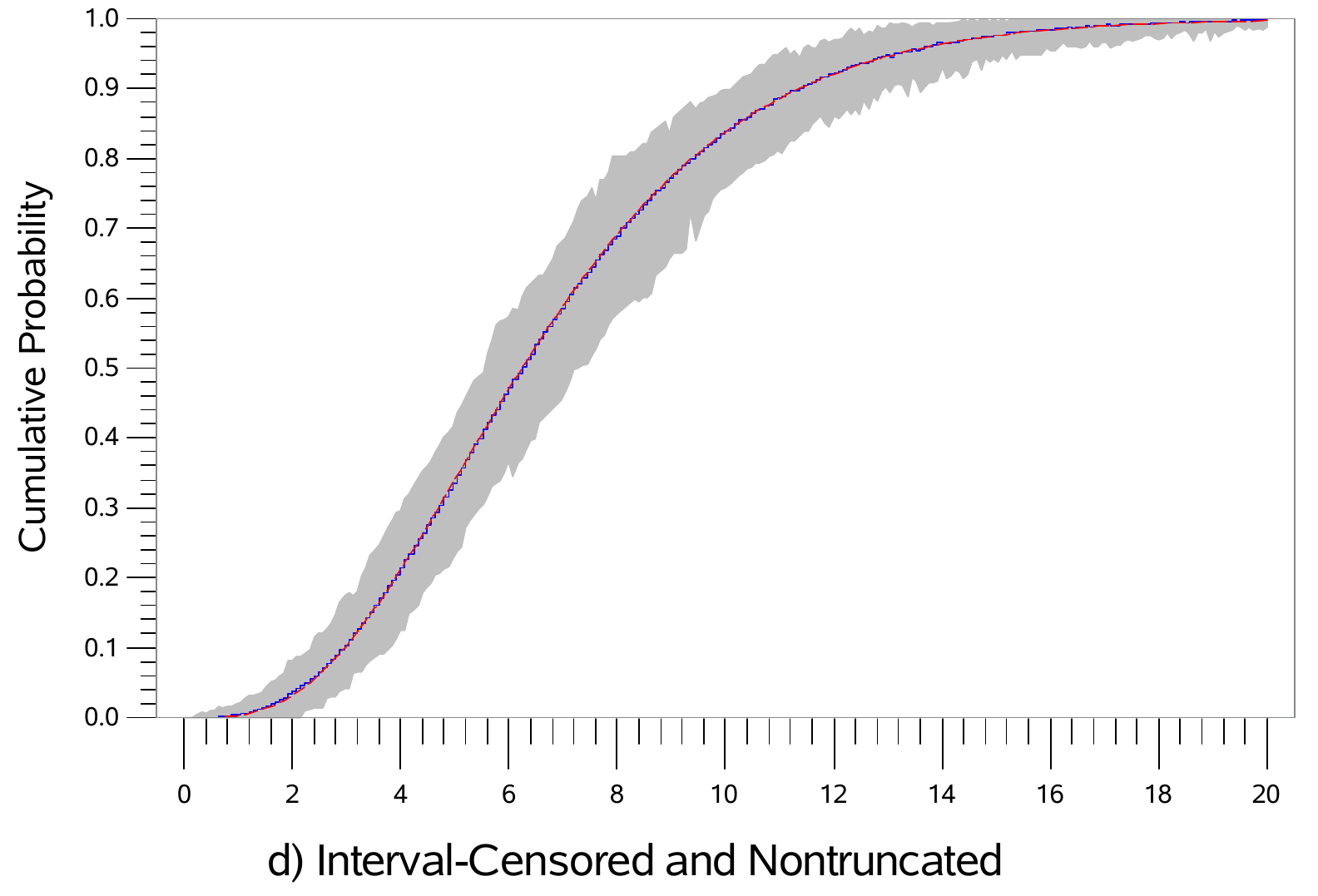} \cr
    \includegraphics[width=0.37\textwidth]{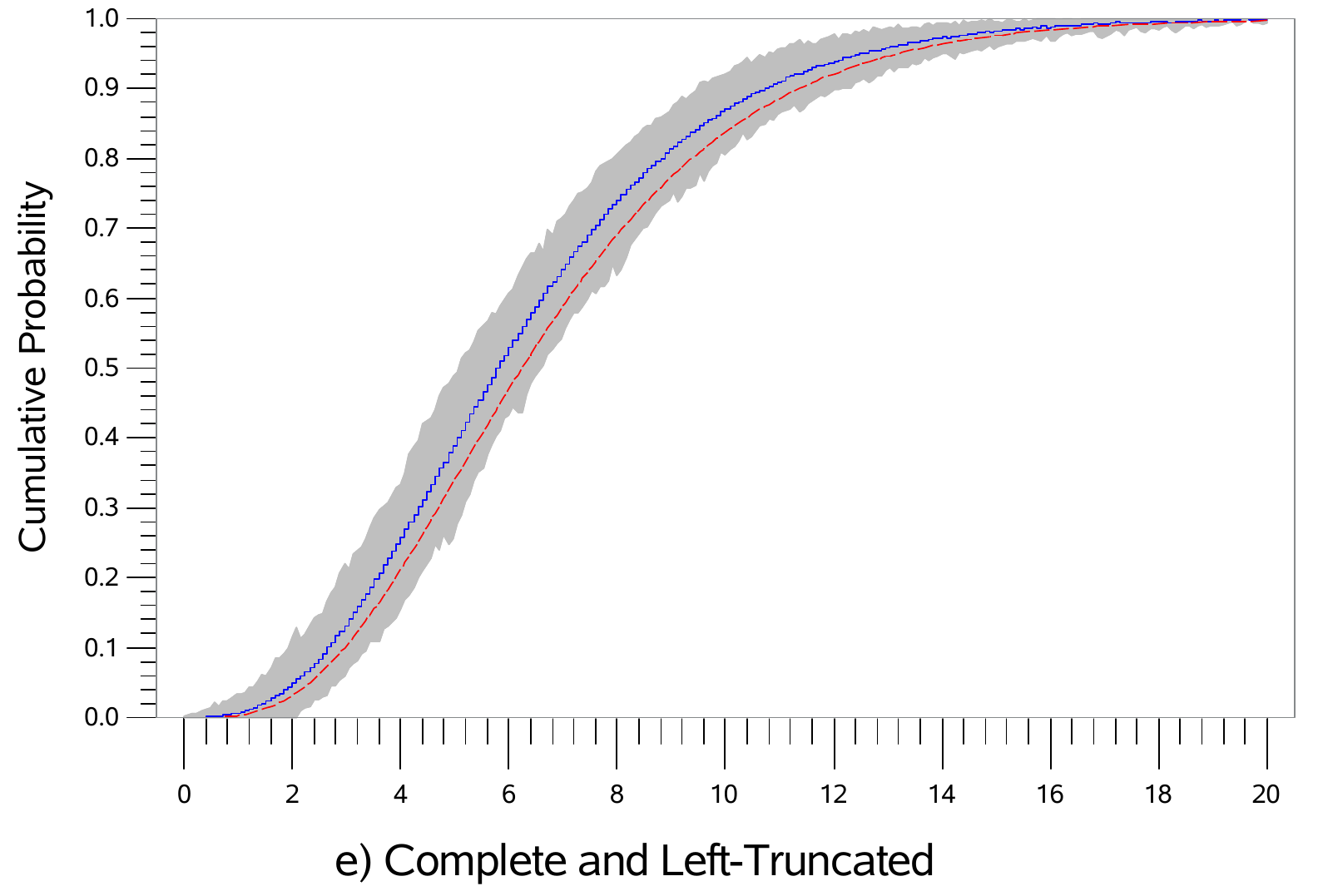} &
    \includegraphics[width=0.37\textwidth]{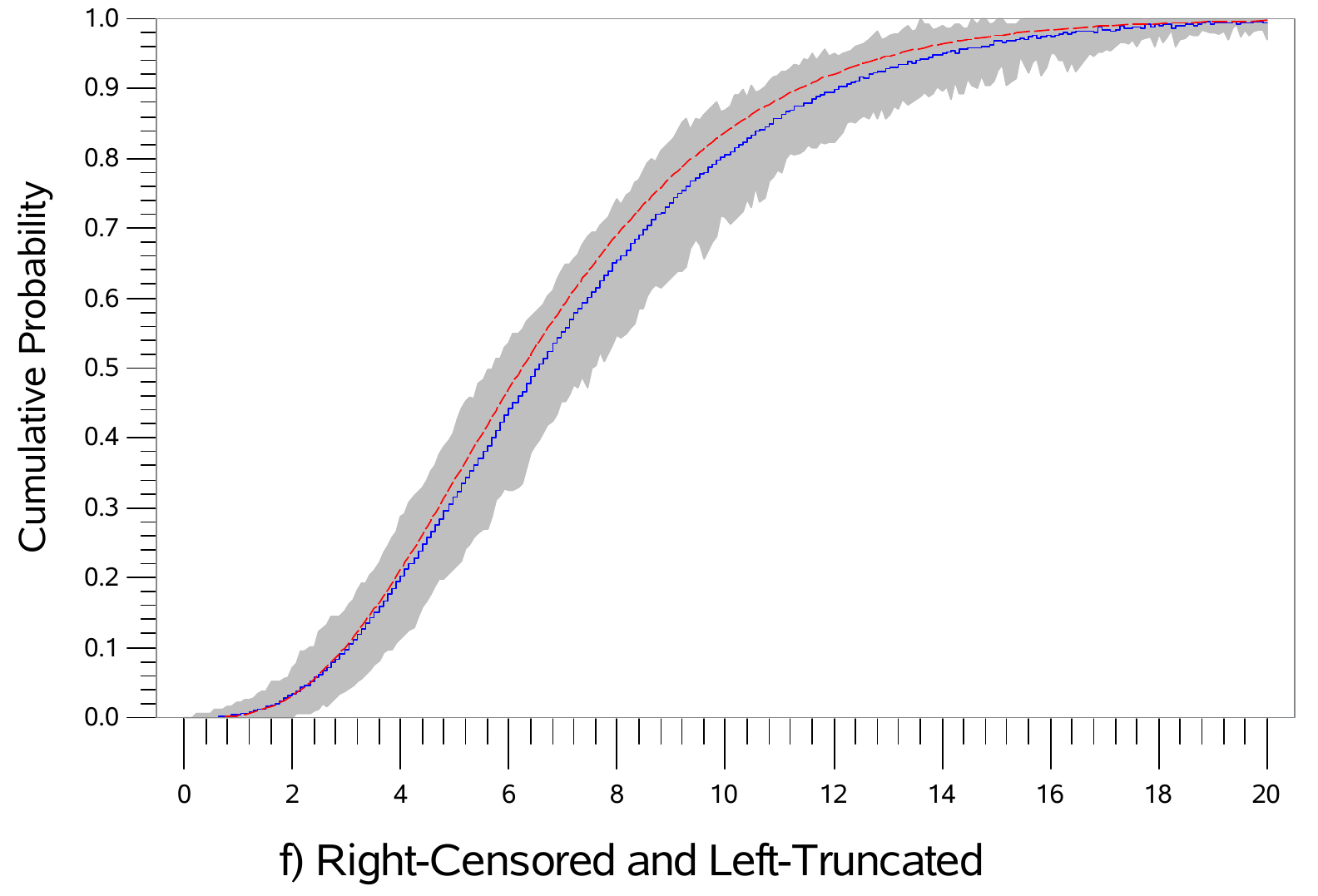} &
    \includegraphics[width=0.37\textwidth]{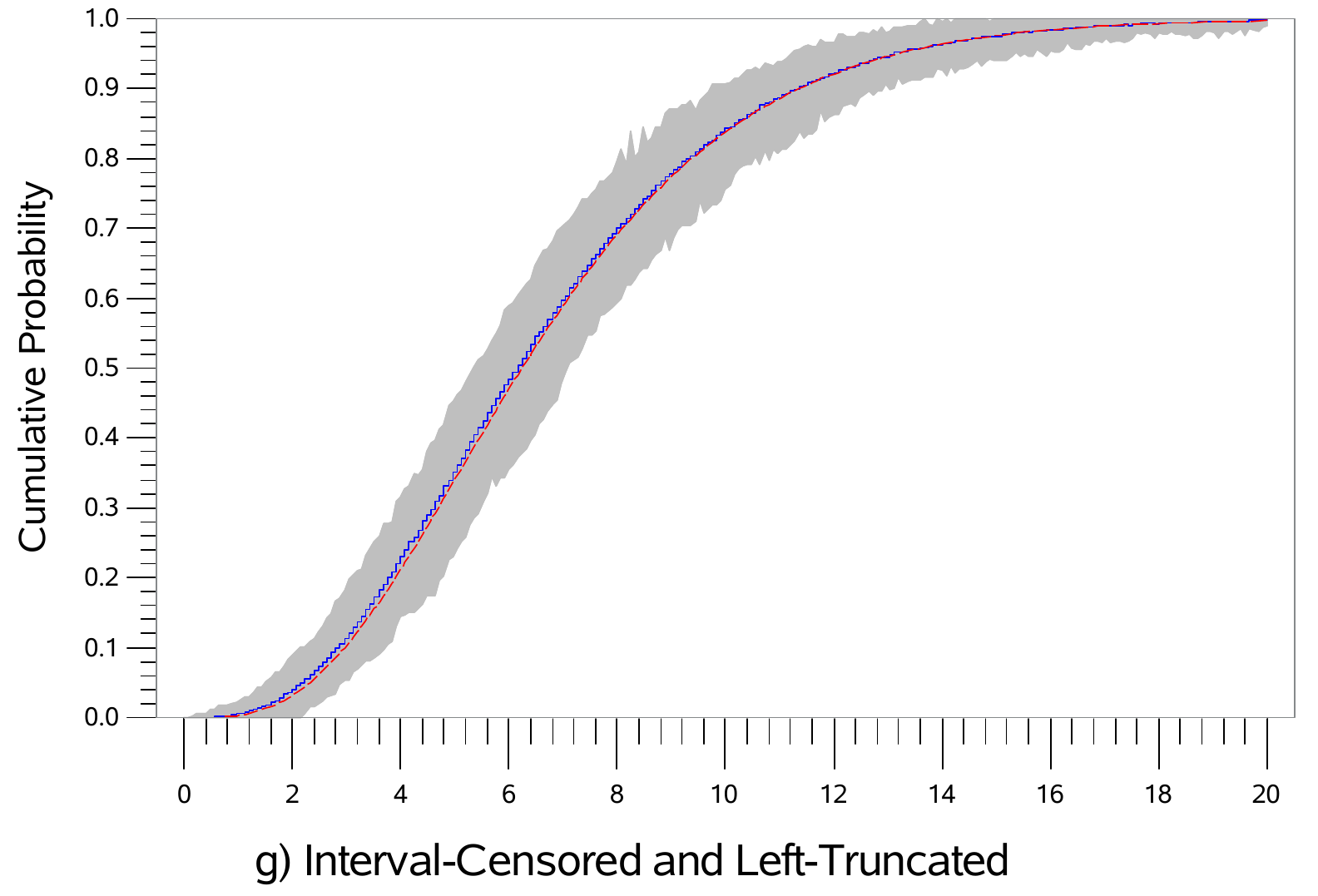} &
    \includegraphics[width=0.37\textwidth]{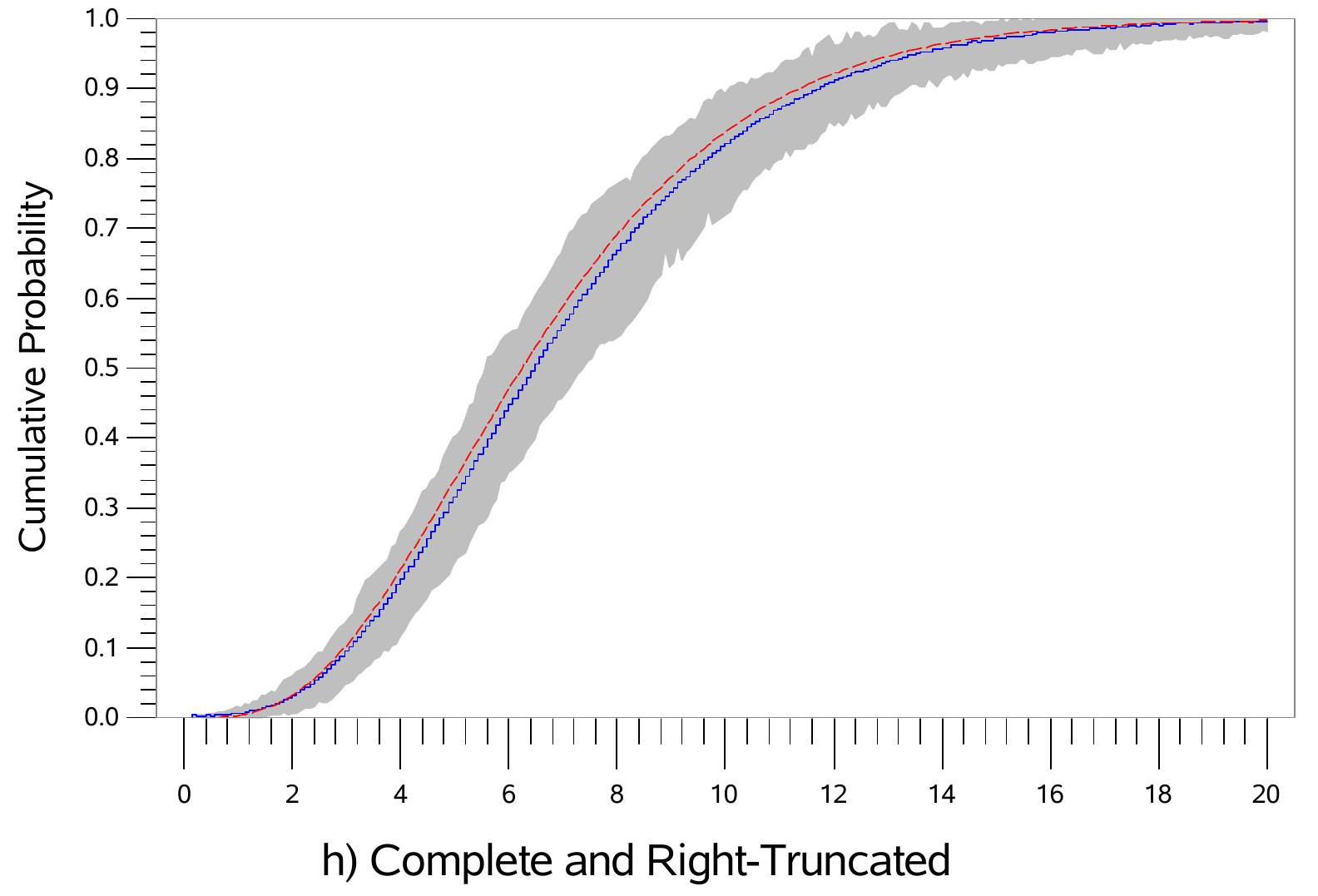} \cr
    \includegraphics[width=0.37\textwidth]{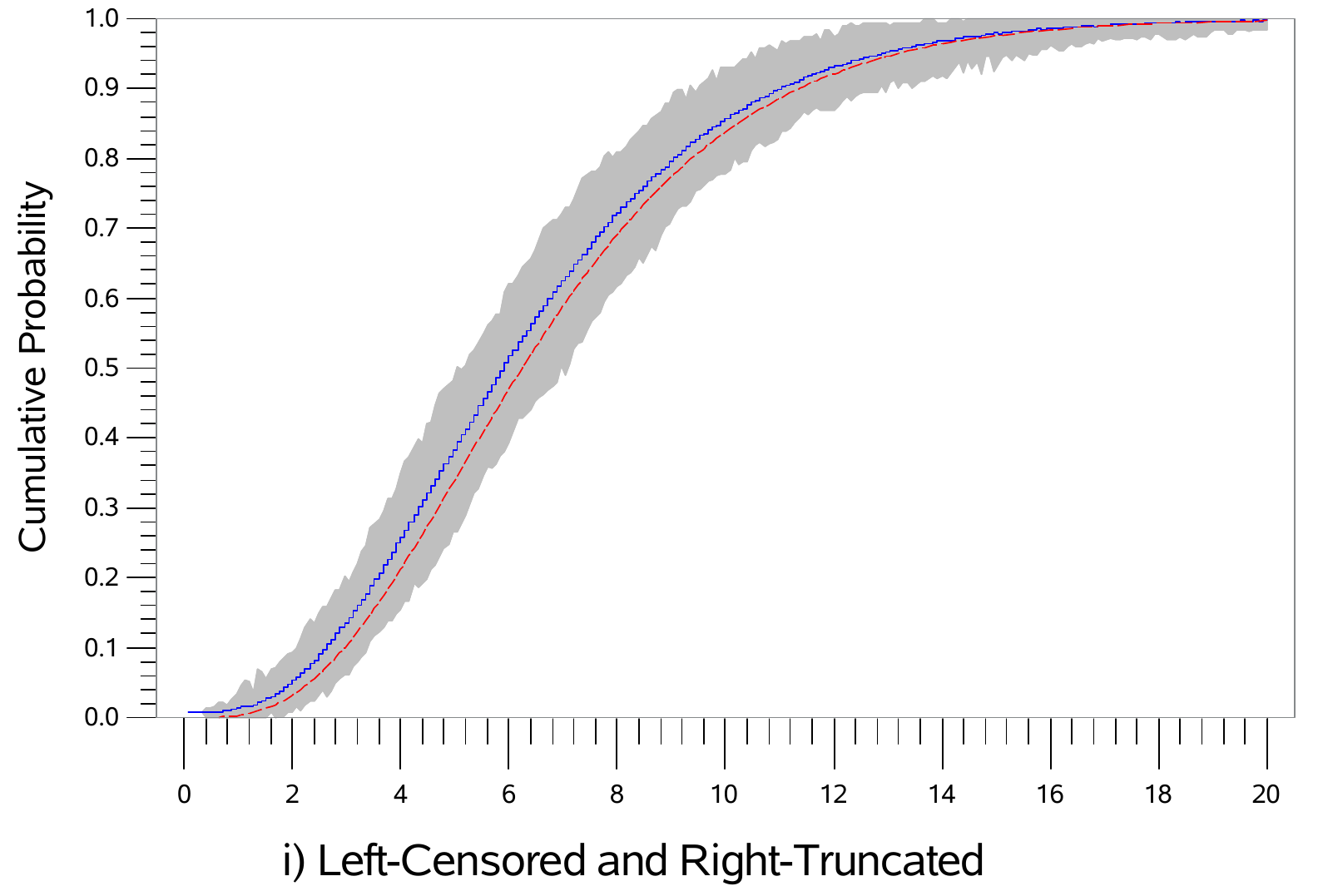} &
    \includegraphics[width=0.37\textwidth]{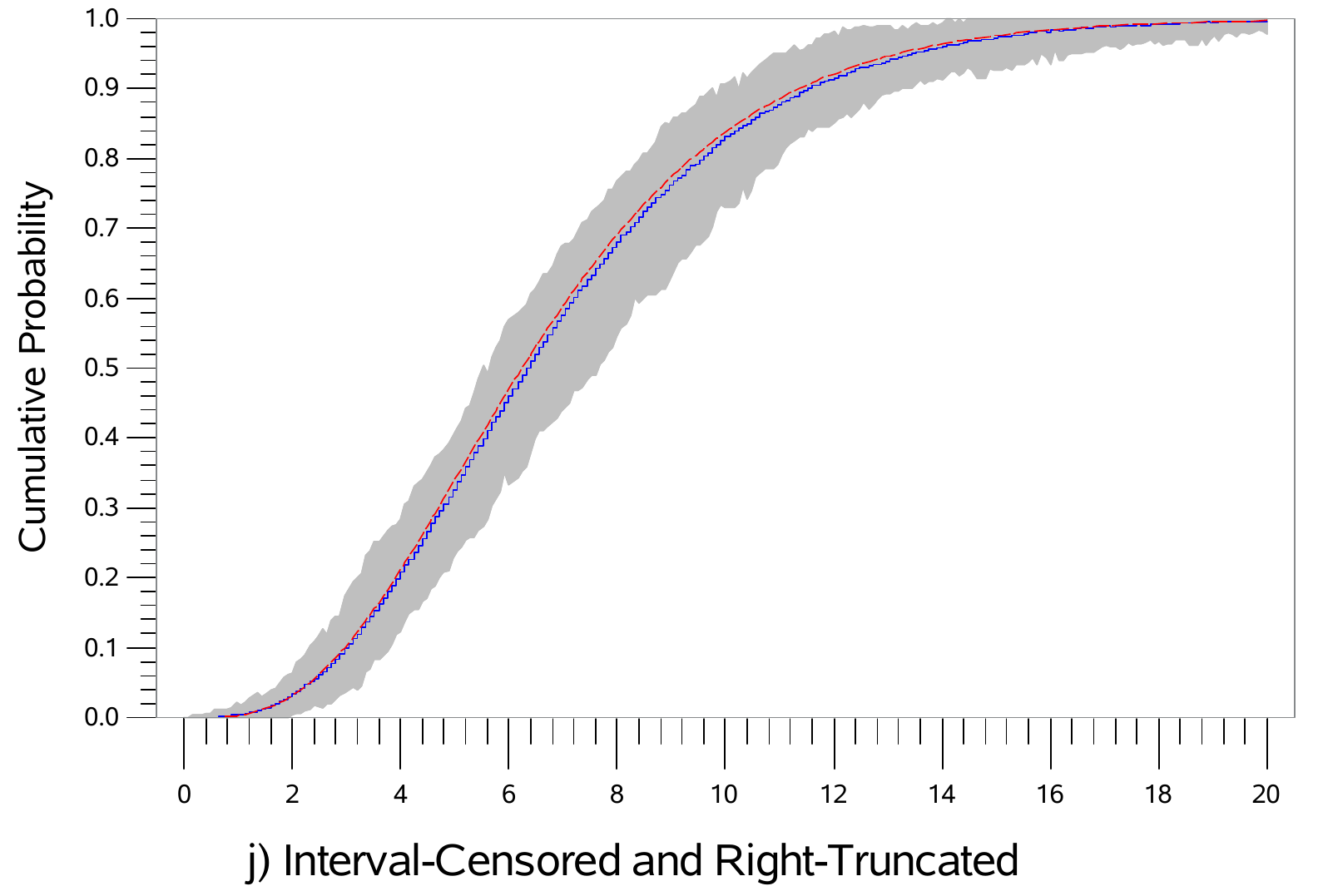} &
    \includegraphics[width=0.37\textwidth]{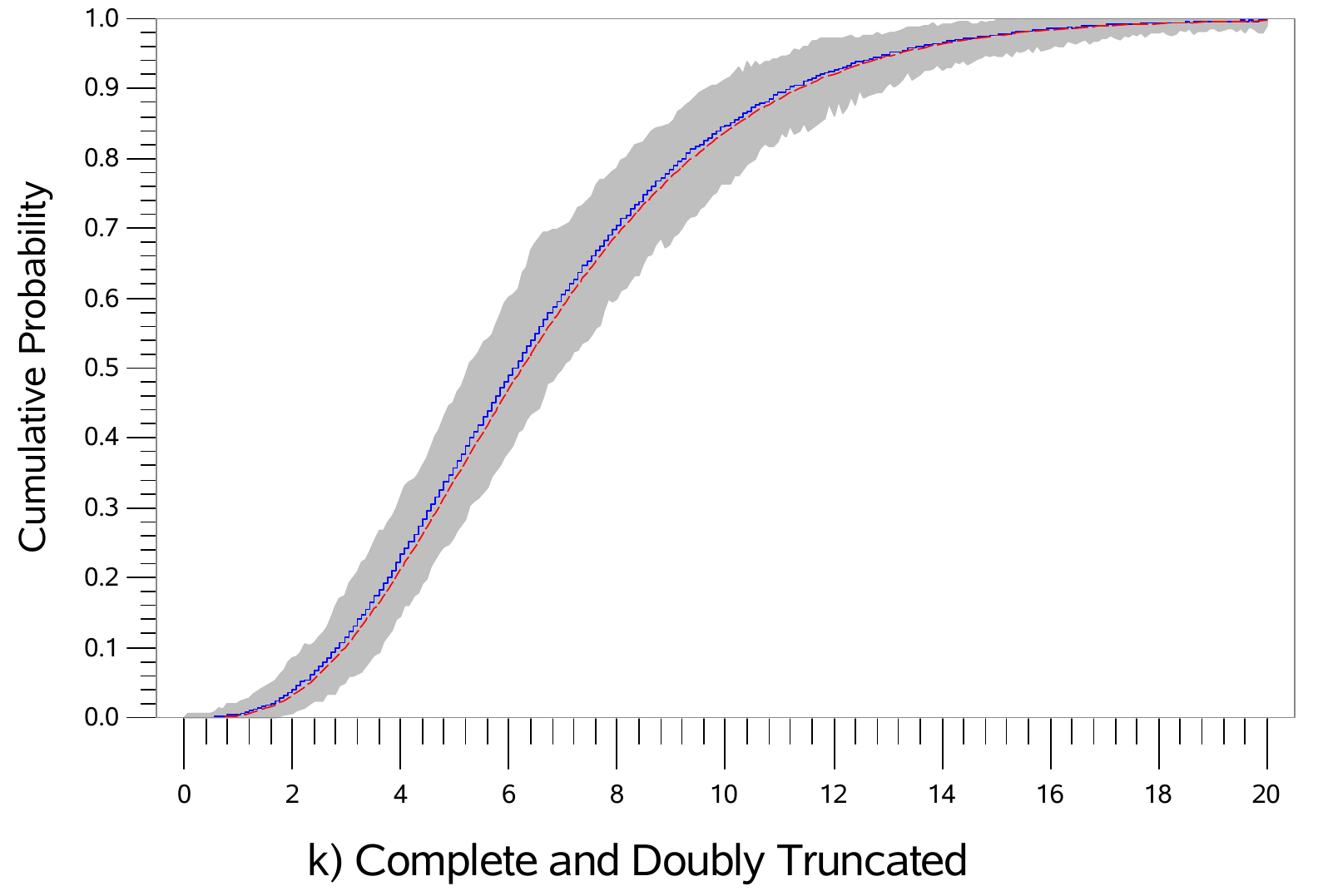} &
    \includegraphics[width=0.37\textwidth]{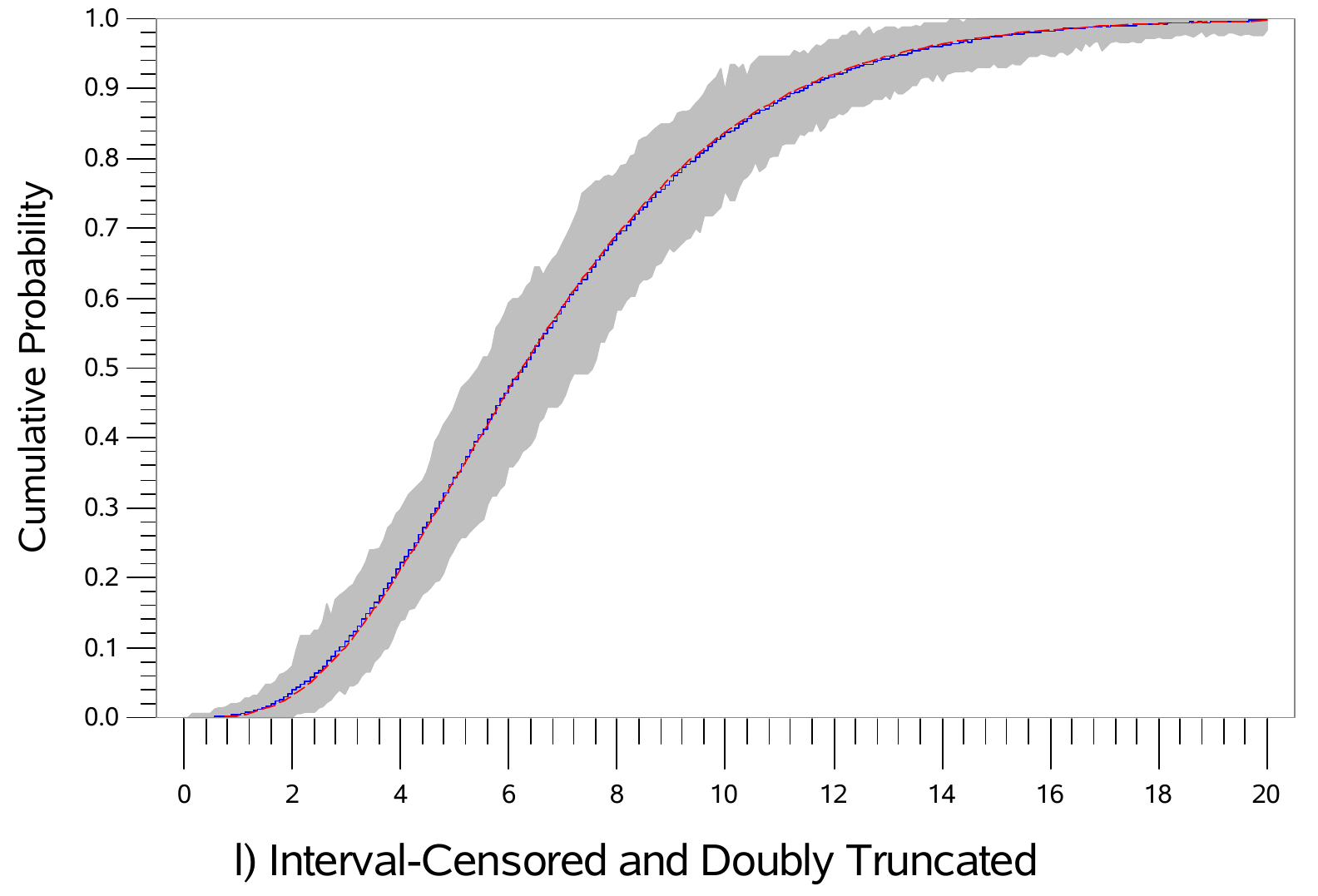}      

    \end{tabular}
    \end{center}
\caption{The mean of qED and 99.9\% confidence bands depending on a sampling scheme
( -~-~-~-~true distribution and  ------~qED)}
    \label{fig:7}
\end{figure*}
\end{landscape}

\paragraph{Example 8.} The effect of sampling schemes on the accuracy of qED for truncated and censored samples is illustrated by Fig.~\ref{fig:7}. A total of 12 types of sampling schemes containing 70\% of complete observations and 30\% of truncated and/or censored observations referred to only one type of those considered in Table~\ref{tab:2} were studied.  For each sampling scheme we carried out a series of 1000 simulations of size $n=250$ from a Gamma distribution with the shape parameter $a=4$ and the scale parameter $\lambda =1.7$.
  
It should be noted that of the 12 sampling schemes considered, only 5 are truncated-censored and 3 are truncated ones. 
The remaining schemes (3 schemes are censored and one is complete) are included in the simulation program to ensure that the study is complete. 
The algorithm proposed in the  Baskakov's paper~\cite{Baskakov1996}, is actually used to construct qED based on censored data, and the usual empirical distribution is used for complete data. 
Therefore, estimates based on complete and censored data can serve as a benchmark when comparing the accuracy of truncated and truncated-censored data.

The analysis of Fig.~\ref{fig:7} shows that in the half of truncation and/or censoring schemes considered the distribution function estimate is biased.
For the three schemes (left-censored, left-censored and right-truncated, and right-truncated) it is  overestimated and for the other three (right-censored, right-censored and left-truncated, and left-truncated) it is underestimated. 
In truncated-censored schemes, the direction of a bias in the distribution function estimate is determined by the type of censoring. 
In schemes with interval-censored or bilateral truncated data, the bias in the expected value of qED does not exceed 1\%. 
In all the schemes considered, the true distribution function is within 90\%  confidence band.

Table~\ref{tab:4} contains the mean Chebyshev distance $\textbf{E}[\rho]$ and relative error of the median  $\textbf{E}[\delta]$, and their quantiles  $\rho_{0.01}$, $\rho_{0.99}$, $\delta_{0.01}$, $\delta_{0.99}$, respectively  1\% and 99\%. 
The maximum value of  $\textbf{E}[\rho]=0.086$ and the same-time maximum median  bias equal to $(1-\textbf{E}[\delta]) \cdot 100=-7\%$ is observed in right-censored samples. 
This fact seems counterintuitive, since in this case the qED is a Kaplan-Meier estimator (see Baskakov~\cite{Baskakov1996} for a proof), that is, a widely used estimator with proven “good” asymptotic properties.

A higher place in the ranking of sampling schemes in terms of the accuracy of  distribution functions estimates is held by left-truncated samples (a median  bias of 6.2\%) and right-censored and left-truncated sample with an  bias of $-5.4\%$. 
Obviously, the most accurate estimate is a usual empirical distribution  based on the complete sample. 
A median  bias in this case equals to 0.2\%. 
Then, interval-censored, non-truncated, and double-truncated samples follow with an $-1.0\%$  bias. 
Thus, according this criterion, the complete data are significantly superior to other sampling schemes. 
However, should the Chebyshev distance be considered as a criterion for determining the accuracy of distribution function estimate, the difference between sampling schemes will no longer be so striking (see Table~\ref{tab:4}).

\begin{table}%

\begin{center}
\caption{\label{tab:4} The Chebyshev distance and the relative error of the median, depending on sampling schemes}
\begin{tabular}{lcccccc}
\noalign{\smallskip}\hline\noalign{\smallskip}

& \multicolumn{3}{c}{~~~Chebyshev Distance~~~} & \multicolumn{3}{c}{Relative Error }\\  \cline{2-7}

\raisebox{1.9ex}[0.4cm][0.1cm]{Type of observations~~}
& \multicolumn{1}{c}{${\rho}_{0.01}$}	& \multicolumn{1}{c}{~$\textbf{E}[\rho]$~}	& \multicolumn{1}{c}{${\rho}_{0.99}$}	& \multicolumn{1}{c}{${\delta}_{0.01}$ }	& \multicolumn{1}{c}{~$\textbf{E}[\delta]$~}	& \multicolumn{1}{c}{${\delta}_{0.99}$}	  \\
\noalign{\smallskip}\hline\noalign{\smallskip}
Complete, Nontruncated	            & 0.026	&	0.053	&	0.100	&	0.911	&	0.998	&	1.098	\\
Right-Censored, Nontruncated        & 0.034	&	0.086	&	0.157	&	0.957	&	1.070	&	1.183	\\
Left-Censored, Nontruncated	        & 0.031	&	0.066	&	0.131	&	0.864	&	0.963	&	1.062	\\
Interval-Censored, Nontruncated	    & 0.029	&	0.060	&	0.121	&	0.905	&	1.010	&	1.124	\\
Complete, Left-Truncated	        & 0.030	&	0.078	&	0.138	&	0.850	&	0.938	&	1.036	\\
Right-Censored, Left-Truncated	    & 0.033	&	0.078	&	0.147	&	0.953	&	1.054	&	1.185	\\
Interval-Censored, Left-Truncated	& 0.031	&	0.061	&	0.120	&	0.882	&	0.993	&	1.113	\\
Complete, Right-Truncated	        & 0.028	&	0.066	&	0.127	&	0.943	&	1.044	&	1.155	\\
Left-Censored,Right-Truncated	    & 0.033	&	0.076	&	0.146	&	0.844	&	0.955	&	1.058	\\
Interval-Censored, Right-Truncated	& 0.034	&	0.066	&	0.120	&	0.925	&	1.026	&	1.124	\\
Complete, Doubly-Truncated	    &   0.028	&	0.057	&	0.110	&	0.881	&	0.987	&	1.095	\\
Interval-Censored, Doubly-Truncated	& 0.031	&	0.061	&	0.119	&	0.897	&	1.010	&	1.129	\\

\noalign{\smallskip}\hline
\end{tabular}
\end{center}
\end{table}

\paragraph{Example 9.} The accuracy of the proposed versus alternative estimates of the distribution function was compared for the data from Gamma  ($a=4$; $\lambda =1.7$), Lognormal ($\theta =0$; $\lambda =1.2$), Tweedie ($\rho=4$; $\mu =1$; $\phi =1$) and Weibull ($a=5$; $\lambda =10$) distributions. 
For each probability distribution, we carried out a series of 1000 simulations of size $n=500$. 
The size of the complete sample is approximately equal to $N \approx 1.59 \cdot n$. 
In all cases, the structure of truncated -censored data was the same (see Table~\ref{tab:3}). 
Each sample contained about 10\% of complete, 60\% of censored, and 70\% of truncated observations on average. 

\begin{figure}
\center
  \includegraphics[scale=0.56]{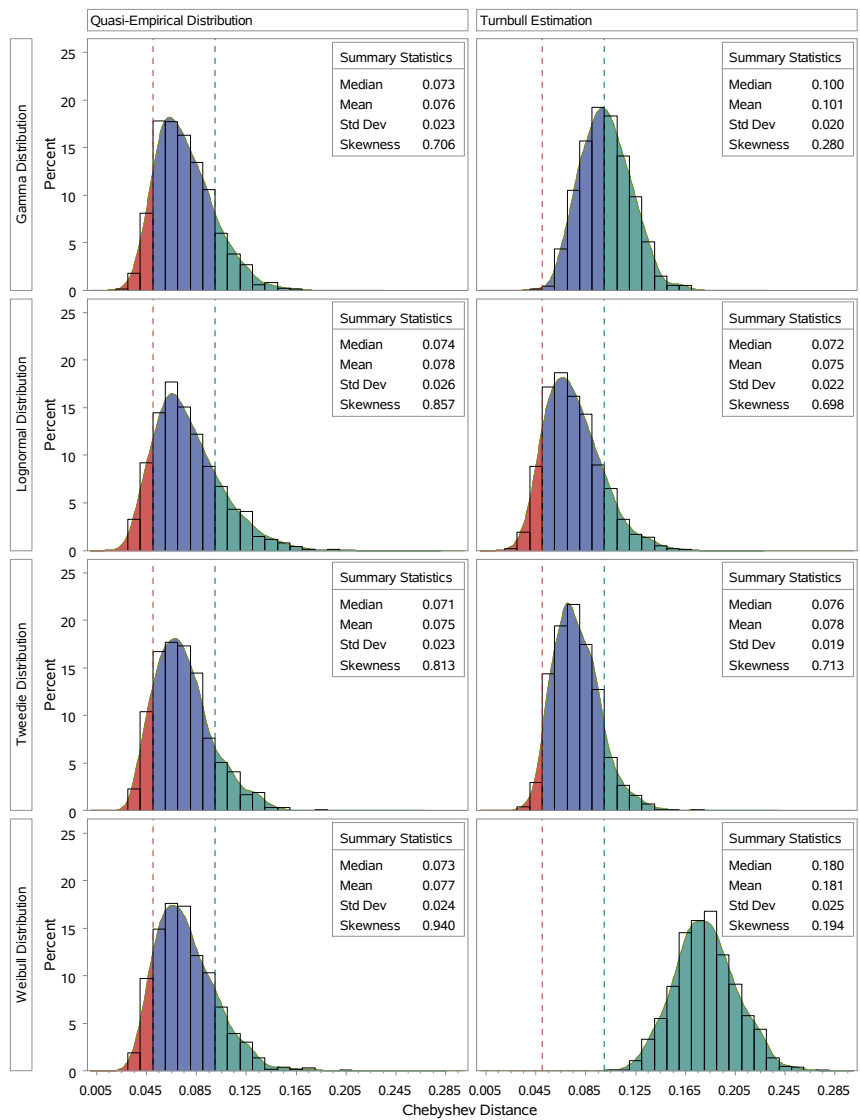}
\caption{Comparing the accuracy of qED and Turnbull estimators for Gamma, Lognormal, Tweedie and Weibull distributions}
\label{fig:8}       % Give a unique label
\end{figure}

\begin{figure}
\center
  \includegraphics[scale=0.56]{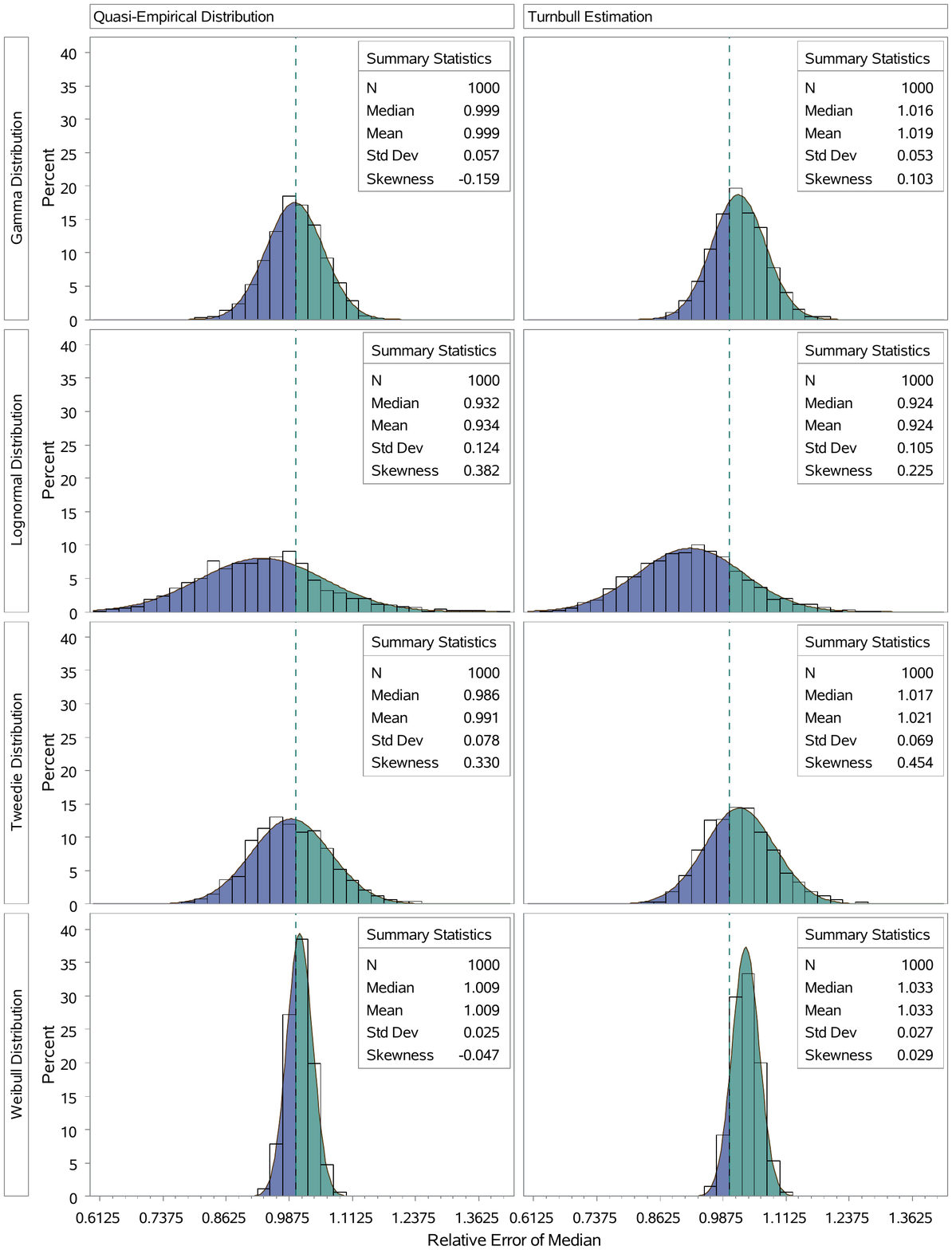}
\caption{Comparing the accuracy of median estimate of qED and Turnbull estimator for Gamma, Lognormal, Tweedie and Weibull distributions}
\label{fig:9}       % Give a unique label
\end{figure}

The qED was estimated by the formula (\ref{equ:20}). As an alternative the Turnbull method~\cite{Turnbull1976},  modified by Frydman~\cite{Frydman1994}, was used for truncated samples (hereinafter referred to as Turnbull estimator). As a criterion for determining the accuracy of distribution function estimation we used the Chebyshev distance  $\rho$ and the accuracy of the median --- the relative error $\delta$. The histograms of corresponding statistics obtained are presented in Fig.~\ref{fig:8} and Fig.~\ref{fig:9}. 

The qED accuracy is almost independent of the type of probability distribution, unlike in the Turnbull estimator. 
A comparison of the median, mean, and standard deviation of distribution $\rho$ shows that for qED, the difference between the minimum and maximum values of these statistics does not exceed 4.1\%, 3.8\%, and 11.5\%, respectively. 
For the Turnbull estimator, the similar indicators are 60.0\%, 58.6\%, and 13.6\%. 
In absolute terms, the mean and median deviation of qED and  true distributions are less than those for the Turnbull estimator for all the distributions considered except for the lognormal one (in the latter case, the difference is 0.003 or 3.8\%). 
At the same time, the standard deviation is less for the Turnbull estimator,  which may be a consequence of a systematic bias which is more pronounced than that for qED estimator. 
These findings are confirmed by a visual comparison of diagrams shown in Fig.~\ref{fig:8}. To make the Figure more informative, two reference lines were added, corresponding to the maximum deviation of the estimate from the true distribution by 0.05 and 0.10. 
The maximum deviation of the Turnbull estimator for a Gamma distribution in almost 50\% of cases exceeds the value of 0.1, and for a Weibull distribution, a similar statement is true in 100\% of cases.

The results of comparing the estimate of median accuracy are presented in Fig.~\ref{fig:9}. One may notice that the conclusions made on the basis of these calculations, partly differ from those previously obtained based on Fig. 8 data analysis. Indeed, the accuracy of the qED median is higher than that of the Turnbull estimator for all the distributions considered (including the lognormal one), and the standard deviation of the estimate of the qED median for a Weibull distribution is lower. At the same time, these calculations confirm the conclusion of a higher bias in case of employing the Turnbull estimator. The maximum bias of 7.6\% and the standard deviation of 0.105 are observed when evaluating data from lognormal distribution. In this case, the mean of Chebyshev distance is minimal (see Fig.~\ref{fig:8}). It should be noted that in all  cases considered, distribution of the median's relative error is satisfactorily approximated by normal distribution.

The results of simulation presented in this Section make no claim for completeness, but allow, in our opinion,to make the conclusion on the effectiveness of the proposed distribution function estimator based on truncated-censored data.

\section{Estimation of multivariate distribution for truncated and censored data}
\label{sec:7}

%В качестве примера построения непараметрической оценки многомерной функции распределения по усеченно-цензурированным данным рассмотрим оценку совместного распределения возраста и стажа работника на предприятии. Поставщиком соответствующей информации обычно выступает отдел кадров предприятия, схема сбора информации и структура получаемых таким образом данных подробно рассмотрены в Примере 5.

%В табл. \ref{tab:5} и на рис. \ref{fig:10} приведена оценка совместной функции распределения возраста и стажа работников крупного предприятия энергетической отрасли по данным отдела кадров за период с 01.01.2010 по 31.12.2013 годы. Объем выборки составил 26~839 человек, включая 4~328 работников, уволившихся по собственному желанию, и 231 работников, умерших за этот период. Взамен выбывших на предприятие было принято 5~129 новых работников. Таким образом, степень усечения данных составила около 0.81, а степень цензурирования – 0.83. 

Let us consider the estimation of the joint distribution of age and years of service of an employee as an example of nonparametric estimation of a multivariate distribution function based on truncated and censored data. HR department is usually a supplier of the relevant information; the collection scheme and the structure of data obtained were discussed in details in Example 6. 

\begin{figure}
\center
  \includegraphics[scale=0.8]{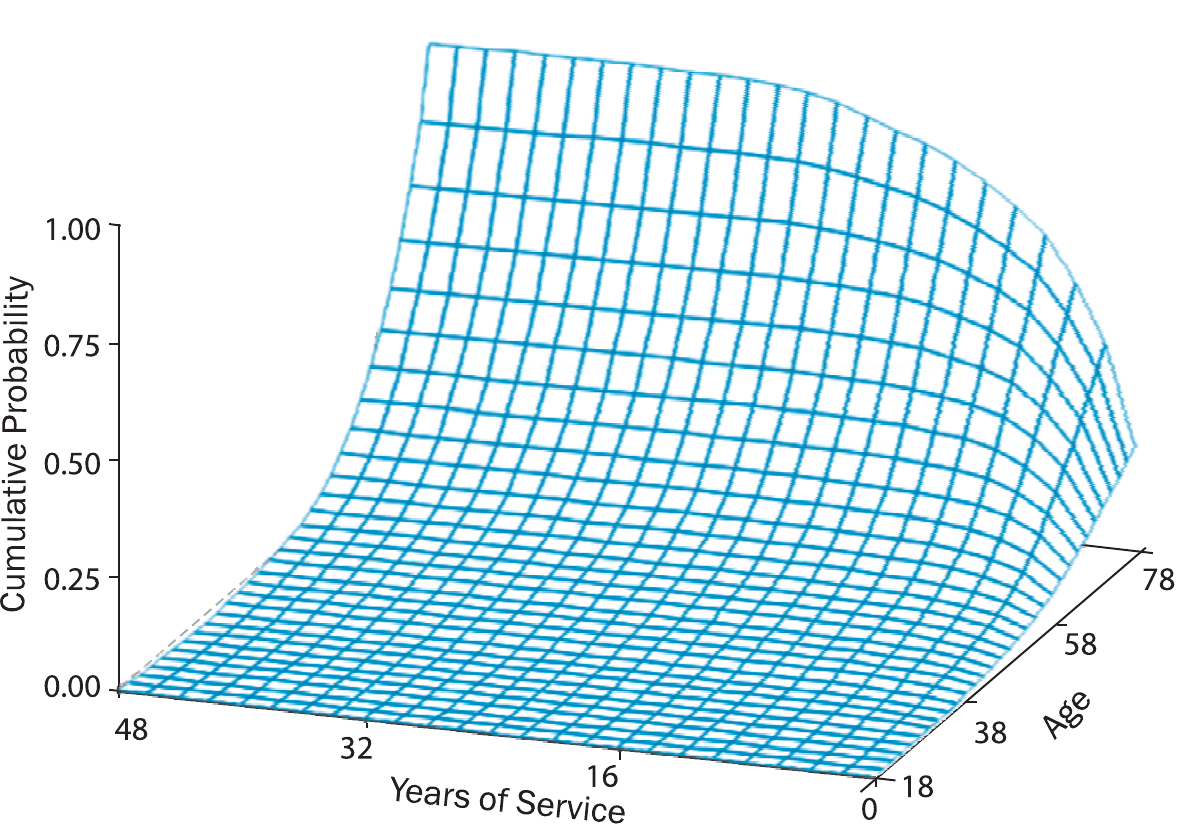}
\caption{The estimation of bivariate distribution of employees' age and years of service }
\label{fig:10}       % Give a unique label
\end{figure}

Table \ref{tab:5}  and Fig. \ref{fig:10}  represent the estimation of the joint function of age and length of service distribution of a large energy enterprise's employees’ based on the data of HR department from 01.01.2010 to 31.12.2013. The sample volume was 26~839, including 4~328 resigned employees and 231 employees who died during this period. At the same time, the company recruited 5 129 new employees. Therefore, the degree of data truncation was approximately 0.81, and the degree of censoring was around 0.83.

\begin{landscape}
\begin{table}
\caption{\label{tab:5} The estimation of the bivariate distribution function of the employees' age and years of service}
\begin{tabular}{ccccccccccccccc}
\hline\noalign{\smallskip}
& \multicolumn{14}{c}{Years of Service} \\ [0.2cm]
\cline{2-15}
\raisebox{1.9ex}[0.4cm][0.1cm]{Age~~~}
& \multicolumn{1}{c}{0}	& \multicolumn{1}{c}{4}	& \multicolumn{1}{c}{8}	& \multicolumn{1}{c}{12}	& \multicolumn{1}{c}{16}	& \multicolumn{1}{c}{20}	& \multicolumn{1}{c}{24}	& \multicolumn{1}{c}{28}	& \multicolumn{1}{c}{32}	& \multicolumn{1}{c}{36}	& \multicolumn{1}{c}{40}	& \multicolumn{1}{c}{44}	& \multicolumn{1}{c}{48}	& \multicolumn{1}{c}{52}\\
\noalign{\smallskip}\hline\noalign{\smallskip}
%18&	0.0000&	0.0000&	0.0000&	0.0000&	0.0000&	0.0000&	0.0000&	0.0000&	0.0000&	0.0000&	0.0000&	0.0000&	0.0000&	0.0000\\
20&	0.0001&	0.0001&	0.0001&	0.0001&	0.0001&	0.0001&	0.0001&	0.0001&	0.0001&	0.0001&	0.0001&	0.0001&	0.0001&	0.0001\\
22&	0.0003&	0.0009&	0.0009&	0.0009&	0.0009&	0.0009&	0.0009&	0.0009&	0.0009&	0.0009&	0.0009&	0.0009&	0.0009&	0.0009\\
24&	0.0006&	0.0027&	0.0045&	0.0045&	0.0045&	0.0045&	0.0045&	0.0045&	0.0045&	0.0045&	0.0045&	0.0045&	0.0045&	0.0045\\
26&	0.0011&	0.0058&	0.0077&	0.0077&	0.0077&	0.0077&	0.0077&	0.0077&	0.0077&	0.0077&	0.0077&	0.0077&	0.0077&	0.0077\\
28&	0.0017&	0.0072&	0.0092&	0.0092&	0.0092&	0.0092&	0.0092&	0.0092&	0.0092&	0.0092&	0.0092&	0.0092&	0.0092&	0.0092\\
30&	0.0025&	0.0104&	0.0126&	0.0126&	0.0126&	0.0126&	0.0126&	0.0126&	0.0126&	0.0126&	0.0126&	0.0126&	0.0126&	0.0126\\
32&	0.0034&	0.0120&	0.0145&	0.0146&	0.0146&	0.0146&	0.0146&	0.0146&	0.0146&	0.0146&	0.0146&	0.0146&	0.0146&	0.0146\\
34&	0.0045&	0.0139&	0.0169&	0.0172&	0.0172&	0.0172&	0.0172&	0.0172&	0.0172&	0.0172&	0.0172&	0.0172&	0.0172&	0.0172\\
36&	0.0060&	0.0177&	0.0211&	0.0216&	0.0222&	0.0222&	0.0222&	0.0222&	0.0222&	0.0222&	0.0222&	0.0222&	0.0222&	0.0222\\
38&	0.0077&	0.0205&	0.0253&	0.0261&	0.0268&	0.0268&	0.0268&	0.0268&	0.0268&	0.0268&	0.0268&	0.0268&	0.0268&	0.0268\\
40&	0.0097&	0.0256&	0.0310&	0.0321&	0.0329&	0.0329&	0.0329&	0.0329&	0.0329&	0.0329&	0.0329&	0.0329&	0.0329&	0.0329\\
42&	0.0122&	0.0300&	0.0370&	0.0392&	0.0412&	0.0412&	0.0412&	0.0412&	0.0412&	0.0412&	0.0412&	0.0412&	0.0412&	0.0412\\
44&	0.0151&	0.0353&	0.0431&	0.0457&	0.0478&	0.0481&	0.0481&	0.0481&	0.0481&	0.0481&	0.0481&	0.0481&	0.0481&	0.0481\\
46&	0.0187&	0.0443&	0.0531&	0.0561&	0.0583&	0.0593&	0.0597&	0.0597&	0.0597&	0.0597&	0.0597&	0.0597&	0.0597&	0.0597\\
48&	0.0227&	0.0516&	0.0617&	0.0651&	0.0674&	0.0689&	0.0693&	0.0693&	0.0693&	0.0693&	0.0693&	0.0693&	0.0693&	0.0693\\
50&	0.0275&	0.0611&	0.0723&	0.0761&	0.0785&	0.0803&	0.0808&	0.0808&	0.0809&	0.0809&	0.0809&	0.0809&	0.0809&	0.0809\\
52&	0.0334&	0.0714&	0.0840&	0.0893&	0.0918&	0.0937&	0.0943&	0.0943&	0.0944&	0.0944&	0.0944&	0.0944&	0.0944&	0.0944\\
54&	0.0405&	0.0853&	0.0998&	0.1055&	0.1084&	0.1103&	0.1109&	0.1109&	0.1110&	0.1110&	0.1110&	0.1110&	0.1110&	0.1110\\
56&	0.0493&	0.1035&	0.1201&	0.1269&	0.1307&	0.1329&	0.1335&	0.1335&	0.1336&	0.1336&	0.1336&	0.1336&	0.1336&	0.1336\\
58&	0.0600&	0.1233&	0.1421&	0.1495&	0.1536&	0.1559&	0.1565&	0.1565&	0.1567&	0.1567&	0.1567&	0.1567&	0.1567&	0.1567\\
60&	0.0729&	0.1498&	0.1728&	0.1809&	0.1855&	0.1881&	0.1889&	0.1889&	0.1891&	0.1891&	0.1891&	0.1891&	0.1891&	0.1891\\
62&	0.0879&	0.1797&	0.2072&	0.2164&	0.2213&	0.2241&	0.2250&	0.2250&	0.2253&	0.2253&	0.2253&	0.2253&	0.2253&	0.2253\\
64&	0.1044&	0.2169&	0.2517&	0.2629&	0.2687&	0.2719&	0.2728&	0.2728&	0.2731&	0.2731&	0.2731&	0.2731&	0.2731&	0.2731\\
66&	0.1214&	0.2599&	0.3039&	0.3185&	0.3252&	0.3287&	0.3297&	0.3298&	0.3301&	0.3301&	0.3301&	0.3301&	0.3301&	0.3301\\
68&	0.1391&	0.3078&	0.3642&	0.3845&	0.3931&	0.3975&	0.3988&	0.3989&	0.3992&	0.3992&	0.3992&	0.3992&	0.3992&	0.3992\\
70&	0.1570&	0.3595&	0.4310&	0.4594&	0.4714&	0.4774&	0.4793&	0.4795&	0.4798&	0.4798&	0.4798&	0.4798&	0.4798&	0.4798\\
72&	0.1761&	0.4154&	0.5040&	0.5432&	0.5610&	0.5705&	0.5736&	0.5741&	0.5744&	0.5744&	0.5744&	0.5744&	0.5744&	0.5744\\
74&	0.1955&	0.4746&	0.5826&	0.6354&	0.6617&	0.6783&	0.6841&	0.6853&	0.6858&	0.6860&	0.6860&	0.6860&	0.6860&	0.6860\\
76&	0.2164&	0.5377&	0.6680&	0.7376&	0.7759&	0.8055&	0.8175&	0.8209&	0.8225&	0.8234&	0.8236&	0.8237&	0.8237&	0.8237\\
78&	0.2373&	0.6033&	0.7582&	0.8482&	0.9021&	0.9522&	0.9770&	0.9860&	0.9911&	0.9960&	0.9972&	0.9986&	0.9989&	1.0000\\
%\hline
\noalign{\smallskip}\hline
\end{tabular}
\end{table}

\end{landscape}

\begin{figure}
\center
  \includegraphics[scale=0.7]{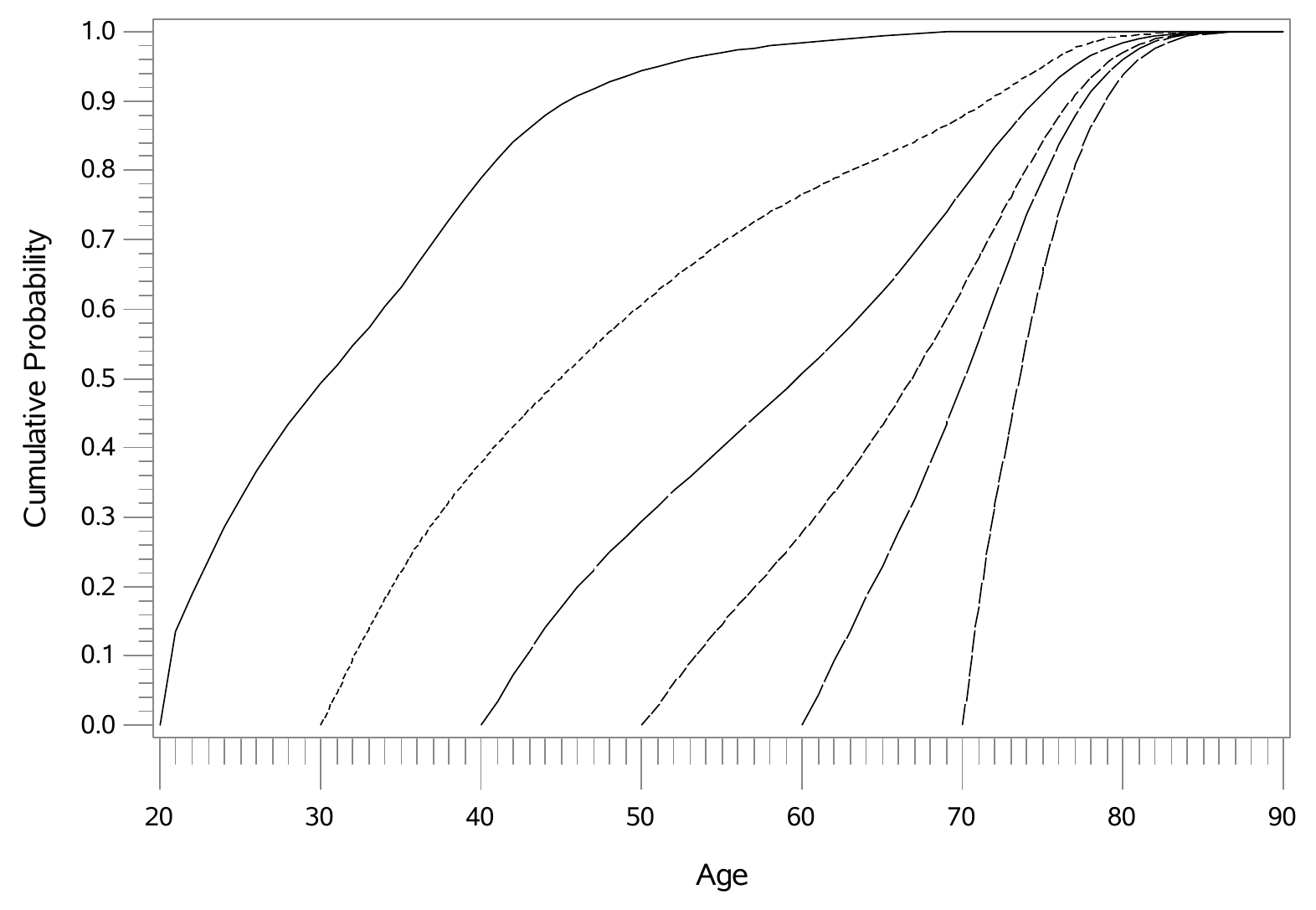}
\caption{The age distribution function of male worker resignation, subject to the age of recruitment, 20, 30, 40, 50, 60, and 70 years (left to right, respectively)}
\label{fig:11}       % Give a unique label
\end{figure}

The obtained estimate of the joint distribution of age and years of service  (see Table~\ref{tab:5} ) allows to effectively calculate the tables of employees' decrement due to resignation, depending on the age of recruitment. The corresponding distribution functions are shown in Fig.~\ref{fig:11}. The calculations made show that the years of service strongly depend on the age of recruitment. Male workers, employed at the age 30 to 45 years, will likely work longer than those employed at both younger and older ages. This appears to be due to the fact that young people have higher labor mobility, while older workers have no time to "earn" long length of service due to natural physical limitations. Note that these patterns have been identified using completely non-parametric methods of statistical data analysis, without any a priori assumptions towards social and economic processes. The results of the calculations obtained are quite consistent with the intuitive perceptions of these processes.   

A detailed solution of the problem discussed in Example 5 is given in Baskakov et al.~\cite{Baskakov2018}. According to \cite{Baskakov2018} the amounts of corporate social liabilities, calculated using the traditional methods of resignation probability assessment or the method proposed in this paper, differ quite significantly. Depending on benefit type, the difference may be over 20\%.

\section{Summary and conclusions}
\label{sec:8}

In this paper we propose a class of non-parametric estimations  of the distribution function based on multivariate truncated-censored data for any sampling scheme.  The shape and size of censoring and truncating sets can be of any type, but should be measurable. In the absence of truncation, the estimate generalizes qED, i.e. non-parametric estimate of the distribution function previously proposed in~\cite{Baskakov1996} for multivariate censored data. The simple and efficient iterative algorithm for constructing a qED on truncated-censored data has been developed and implemented in the SAS/IML environment for different kinds of univariate and some bivariate sample schemes with truncated and censored observations. 

The accuracy of qED and alternative estimates of the cumulative distribution function for some sampling schemes has been compared through simulation studies. The simulation results have proved the satisfactory accuracy of the proposed non-parametric estimates of the distribution function, sufficient to  recommend it for actuarial practice. The algorithm has been tested for many years in the IAAC Group of Companies in the valuation of corporate social liabilities according to IAS 19 Employee Benefits.

It is of interest to further develop the use of qED for new tasks in insurance and pensions, as well as further research of the estimation's properties. We leave this as a future work.

%%%%%%%%%%%%%%%%%%%%%%%%%%%%%%%%%%%%%

%\begin{acknowledgements}
%\end{acknowledgements}



\end{document}